\documentclass{article}
\usepackage{graphicx} 
\usepackage{import}
\usepackage{color,soul}
\usepackage{hyperref}
\usepackage{placeins}
\usepackage{float}

\title{Unsupervised detection of coordinated fake-follower campaigns on social media}
\author{
  Yasser Zouzou\textsuperscript{1}, 
  Onur Varol\textsuperscript{1,2,*}
}

\date{
    \begin{small}
    \textsuperscript{1}Faculty of Engineering and Natural Sciences, Sabanci University \\ \textsuperscript{2}Center of Excellence in Data Analytics, Sabanci University \\
    \textsuperscript{*}Corresponding author: \texttt{onur.varol@sabanciuniv.edu}
    \end{small}
}

\begin{document}

\maketitle

\begin{abstract}
Automated social media accounts, known as bots, are increasingly recognized as key tools for manipulative online activities. 
These activities can stem from coordination among several  accounts and these automated campaigns can manipulate social network structure by following other accounts, amplifying their content, and posting messages to spam online discourse. 
In this study, we present a novel unsupervised detection method designed to target a specific category of malicious accounts designed to manipulate user metrics such as online popularity. 
Our framework identifies anomalous following patterns among all the followers of a social media account. Through the analysis of a large number of accounts on the Twitter platform (rebranded as Twitter after the acquisition of Elon Musk), we demonstrate that irregular following patterns are prevalent and are indicative of automated fake accounts. 
Notably, we find that these detected groups of anomalous followers exhibit consistent behavior across multiple accounts. 
This observation, combined with the computational efficiency of our proposed approach, makes it a valuable tool for investigating large-scale coordinated manipulation campaigns on social media platforms.
\end{abstract}

\section{Introduction}
Twitter was originally established as a personal social networking platform, where users can follow each other and share messages with their followers. In recent years, Twitter has been used by leading politicians and large organizations worldwide for sharing information and news. The transformation of Twitter to a major online venue for sharing information has made it a favorable space for misinformation spreading. 
In order to efficiently spread misinformation, automated accounts, known as \textit{bots}, have been widely used on Twitter \cite{ferrara2016rise,cresci2020decade}. 
Bots are run through Twitter API and are legitimate as long as they openly state on the platform that they are bots \cite{alkulaib2022twitter,yang2019arming}. 
However, bots are also used on social media platforms in a malicious manner to spread misinformation and manipulate user popularity and engagement metrics \cite{bruno2022brexit,varol2020journalists,shao2018spread}. 
Identifying malicious bots is crucial for suspending these accounts and conducting research to understand the role of bots in manipulation campaigns on social media \cite{himelein2021bots,varol2017online,yang2019arming}.
Recently, acquisition of Twitter by Elon Musk also raised concerns about bot prevalence on the platform \cite{varol2023should}.

In recent years, there has been an increasing interest in automatically detecting malicious bots on social media platforms. Cresci thoroughly reviews bot detection approaches for the last decade \cite{cresci2020decade}. 
Bot detection methods can be categorized into supervised and unsupervised methods. Supervised detection methods \cite{sayyadiharikandeh2020detection,liu2023botmoe, ding2023find} require labeled datasets of bot accounts, which constitutes their main weakness. 
Firstly, there is a limited number of labeled datasets of Twitter accounts, and many of the existing ones are outdated and do not capture the evolution of spam accounts and fake followers. 
Secondly, these models struggle to generalize well on unseen bot types in the training dataset\cite{echeverri2018lobo}, but novel approaches can address these limitations with novel machine learning systems\cite{sayyadiharikandeh2020detection}. 
Besides fully supervised approaches, semi-supervised approaches that rely on a small set of labeled bot accounts also exist \cite{jia2017random, mendoza2020bots}. 
In the semi-supervised detection methods, which rely on a network representation of user relationships and interactions, the accounts that are most similar to the labeled bot accounts are considered more suspicious. 
Unsupervised approaches \cite{mazza2019rtbust, mannocci2022mulbot} rely on clustering users based on a set of features and identifying clusters that have suspicious properties or behavior.

\begin{figure}[!htbp]
    \centering
    \includegraphics[width=\textwidth]{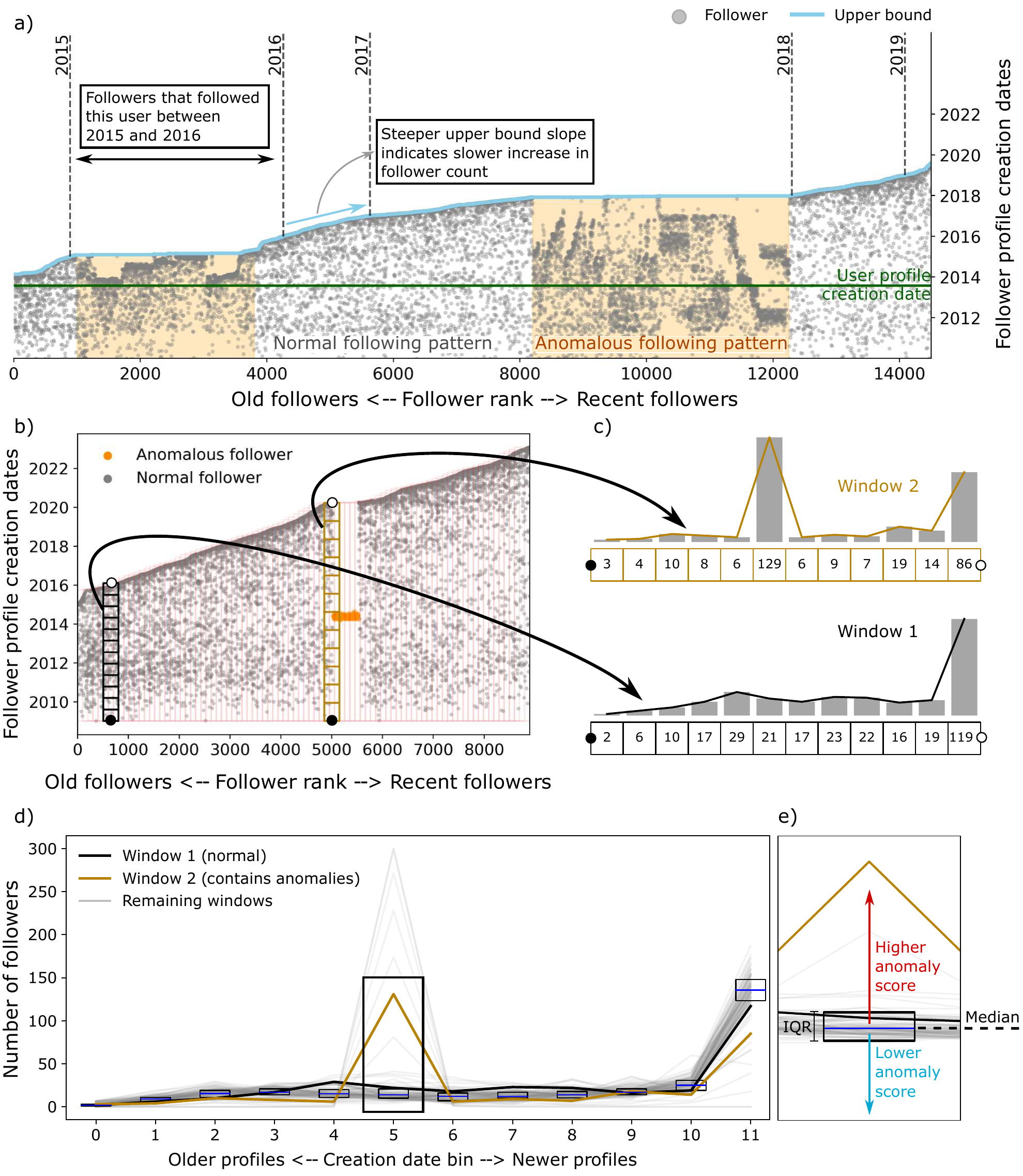}
    \caption{\textbf{Illustration of the follower map and the detection method we introduce (Sliding Histogram)} (a) A follower map with regions of anomalous following patterns highlighted in orange. The vertical dashed lines mark the beginning of each year based on the estimated following times. (b) A follower map with inserted synthetic irregular followers (orange) showing all sliding windows (light gray) with two of them highlighted in black and orange. (c) The histograms corresponding to the two highlighted windows in the follower map. Window 1 only includes normal followers and Window two includes anomalous followers. The numbers are the count of followers that fall within each bin. (d) All histograms plotted together as line plots, with the black and orange lines corresponding to the black and orange windows above. (e) A zoom in on bin No. 5 showing the median and interquartile range (IQR) of all histograms at this bin.}
    \label{fig:follower_map_and_sliding_hist}
\end{figure}

In this study, we propose an unsupervised method to detect a type of anomalous followers that has not been specifically addressed before. Our definition of anomalous followers originates from the findings of a New York Times investigation on fake follower markets\footnote{https://www.nytimes.com/interactive/2018/01/27/technology/social-media-bots.html}, and a subsequent study on journalists on Twitter that demonstrate fake followers were used to increase online popularity and manipulate online perception of journalist accounts \cite{varol2020journalists}. 
To quantify the groups of irregular followers observed in the aforementioned studies, we introduce an instrument that we call \textit{follower map}. A follower map is a graph that plots all the followers of a certain account based on their follow rank (x-axis) and their account creation dates (y-axis). It is important to note that the x-axis corresponds to the order of following and not the exact time of following. Fig. \ref{fig:follower_map_and_sliding_hist}a is a segment of the follower map of a user in our dataset showing the first 15,000 followers of this account. The rising upper bound (blue line) at each rank represents the most recent profile creation date up to the current follower rank. Since we are aware that the time when each follower started following is guaranteed to be after the creation date of all previous followers, we can use the upper bound as a proxy to estimate these follow times. The exact detail of the follow time estimation algorithm can be found in \cite{meeder2011we}. The vertical dashed lines depict the beginning of each year based on the estimated following times. In areas where the follow pattern is normal, followers are evenly distributed along the y-axis, with a slightly denser region near the upper bound. The dense region beside the upper bound is attributed to accounts following this user just after being created. In the anomalous regions of the map (highlighted in orange), we see batches of follower accounts created on similar dates and following the user consecutively. During these anomalous regions, the upper boundary remains horizontal and then returns to its original slope after the anomalous follower batch, suggesting that this group followed the user rapidly. Followers showing this anomalous pattern were found to be automated fake accounts\footnotemark[\value{footnote}] and often had high bot scores \cite{varol2020journalists}. In this study, we propose a method to automatically identify unusual following patterns in the followers of social media accounts. The study is divided into two parts: (i) Finding the most suitable detection method for this type of anomalous followers by testing several approaches on a synthetic dataset (ii) Identifying and analyzing irregular follower patterns in a real Twitter dataset comprising Turkish politician and media outlet accounts \cite{najafi2022secim2023}.

In the first part of this study, we test 4 existing anomaly detection algorithms on features engineered from the follower map and introduce a method specific to this task. In order to apply anomaly detection algorithms, we generate features that capture the local density around followers in the follower map. We expect the anomaly detection methods to assign higher anomaly scores to the groups of unconventional followers since their local density in the follower map is significantly different from the normal followers. On the other hand, our approach, which we refer to as the Sliding Histogram (SH), does not rely on engineered features. Instead, SH compares the distribution of followers in a moving window to the entire distribution of followers in the follower map of a Twitter user (Fig. \ref{fig:follower_map_and_sliding_hist}b-e). 
Since our approach does not rely on handcrafted features, the deviant following patterns that it can identify are not limited to the ones observed earlier. Our method assigns a score for each follower in the follower map of an account, where high scores indicate that the user is part of a group of followers that have an irregular following pattern. We refer to this score as \textit{anomaly score} throughout the paper. Due to the absence of a labeled dataset of anomalous followers as we define them, we first test these methods on a synthetic dataset. The synthetic dataset consists of simulated aberrant followers inserted into the followers of users on the Dribbble platform\footnote{https://dribbble.com/}, which is a social networking platform for digital designers. We choose the Dribbble platform for two reasons: (i) As a professional network for designers, it is less susceptible to having fake followers (ii) Follower data collected from Dribbble includes following times, which allows us to validate the follow-time estimation method \cite{meeder2011we} we use in our analyses in the second part of the study. In total, we test 5 detection methods on a dataset comprising more than 155,000 unique datasets of Dribbble user followers with inserted synthetic anomalous followers.

In the second part of the study, we apply the best-performing detection method on the followers of 1,318 Twitter accounts of Turkish politicians and media outlets taken from the \#Secim2023 dataset \cite{najafi2022secim2023}. Since our approach addresses all the followers of a user, we explore its ability to identify users with irregular followers out of a pool of users, i.e., given a large set of Twitter accounts, which are the accounts that have anomalous followers amongst their followers? Then, we explore the detected anomalous follower accounts and qualitatively verify that they are indeed fake accounts. Finally, we conduct an analysis to explore the coordinated activities of the detected fake accounts.

In summary, we introduce an unsupervised method that can find the previously unexamined deviant following patterns within the followers of a social media account. Our method relies solely on the profile creation dates and ranks of a user's followers, making it adaptable to any platform that offers this data. Furthermore, this approach computes anomaly scores to all of a user's followers instead of looking at individual or small groups of accounts, allowing for the identification of users that purchase (or are targeted by) groups of anomalous followers. We present two use cases for this method: (i) Identifying accounts that have irregular followers from a pool of accounts (ii) Identifying individual automated accounts that engage in coordinated following behavior. Our study demonstrates that these aberrant followers are automated accounts, and in certain instances, they elude existing bot detection methods that rely on inputs.

\FloatBarrier
\section{Methodology}
\subsection{Datasets}


\textbf{Synthetic dataset}: We obtained the profile data of the followers of 2,834 users on the Dribbble platform\footnote{https://dribbble.com/}, a social networking platform for digital designers. The 2,834 collected users have follower counts ranging between 1,000 and 110,000. The distribution of the follower counts and data preparation details are given in Appendix \ref{A:data_details}. 
We insert two types of synthetic anomalous followers that we derive from the previously observed anomalous follow patterns \cite{varol2020journalists}. Type 1 followers represent a batch of follower accounts that were created in a limited range of time and followed the user consecutively. Type 2 followers represent a batch of follower accounts that followed the user consecutively and almost immediately after being created. By varying the count of synthetic followers, the spread of these followers, and their combinations, we generated 55 permutations of the original dataset, resulting in a total of $55 \times 2834$ synthetic datasets, each having distinct synthetic anomalous followers. Further description of the parameters used to generate these synthetic followers is provided in Appendix \ref{A:synthetic_data_generation}.

\textbf{Twitter dataset}: We use the Twitter accounts of 1,318 Turkish politicians and media outlets from the \#Secim2023 dataset \cite{najafi2022secim2023} to experiment with our anomalous follower detection approach. Although this is an unlabeled dataset, it allows us to explore the types of anomalous followers that our approach is capable of capturing in real Twitter data. The number of followers of these accounts ranges between 1,000 and 20 million. The distribution of the follower counts and data preparation details are given in Appendix \ref{A:data_details}

\subsection{Anomalous follower detection}
In order to apply the unsupervised anomaly detection algorithms, we generate features from the follower map that can help isolate the anomalous followers in the feature space. 
The anomalous following patterns we are interested in detecting consist of followers with similar profile creation dates following an account consecutively. This translates to dense regions in the follower map. Therefore, we use features that describe the local density around followers in the follower map. We use features to describe the position of a follower in the follower map to prevent mislabeling dense regions that are typically dense in normal follower maps, such as those near earlier ranks and around the upper bound. The features are described in detail in Appendix \ref{A:feature_engineering}.

We evaluate 4 unsupervised anomaly detection algorithms using the engineered features: (1) Isolation Forest \cite{liu2008isolation} (2) Local Outlier Factor (LOF) \cite{breunig2000lof} (3) Empirical-Cumulative-distribution-based Outlier Detection (ECOD) \cite{li2022ecod} (4) Gen2Out \cite{lee2021gen}. 
In addition, we design a task-specific approach, Sliding Histogram (SH), which does not rely on engineered features and can be applicable to any other social media data if the follower rank and account creation times are available for the analysis.

\subsubsection{Isolation Forest}
In the isolation forest algorithm \cite{liu2008isolation}, a forest of decision trees with random splits is grown, and higher anomaly scores are given to points that have a shorter average path in the forest. The path length is the number of splits from the root node required to isolate a data point in a leaf node. This definition of anomaly score is based on the fact that anomalies, by definition, are "few and different". Therefore, by randomly splitting nodes in a decision tree, we expect anomalies to be isolated earlier than normal points since they reside in sparser areas of the feature space. Isolation forest trees are created using sub-samples of the dataset to avoid two common problems in anomaly detection: swamping and masking. The isolation forest algorithm requires two main hyperparameters: number of trees in the forest and sub-sample size. In our experiment, we use 200 trees and a sub-sample size of 256, which is the size recommended by the authors

\subsubsection{Local Outlier Factor}
The Local Outlier Factor algorithm 
 (LOF) \cite{breunig2000lof}, is designed to detect local outliers, i.e., points that lie in areas with less density than that of the nearest cluster of points. A point is assigned a high anomaly score if the average distance between this point and its nearest neighbors is greater than the average distance between its nearest neighbors and their nearest neighbors. The main hyperparameter in this algorithm is the number of nearest neighbors to be considered (\textit{MinPts}). Since we are dealing with groups of anomalous followers, we expect them to be clustered together in the feature space. Thus, \textit{MinPts} should be set to a value greater than the number of anomalies in a group of anomalous followers. Otherwise, this cluster of anomalies would be assigned low anomaly scores since all the nearest neighbors would be inside the same cluster. However, the fact that we do not have prior information about the number of anomalies that we expect to see in one group makes it hard to choose the value of \textit{MinPts}. In our experiment, we set \textit{MinPts} to 3\% of the total number of followers of each user. Although users may have an anomaly ratio greater than 3\% in their followers, larger values of \textit{MinPts} result in prohibitive run times and memory usage for users with a large number of followers.

\subsubsection{ECOD}
The Empirical-Cumulative-distribution-based Outlier Detection (ECOD) method assigns high outlier scores to data points that have a low tail probability under the joint cumulative distribution function (CDF) of the data \cite{li2022ecod}. The joint CDF is estimated by assuming that the dimensions (features) of the data are independent. Thus, the product of the univariate empirical CDFs (ECDF) of all dimensions is used as an estimate of the joint CDF. Data points that have extreme feature values, based on the distribution of the corresponding feature, receive high outlier scores. This method does not require any hyperparameter tuning and is computationally efficient. However, due to the independence assumption, the interactions between features are not considered in this method.

\subsubsection{Gen2Out}
The Gen2Out method relies on the same concept of the IF method, i.e., an anomalous point tends to have a shorter average path from the root node to its leaf node in a forest of random decision trees, referred to as \textit{AtomTrees} in this study \cite{lee2021gen}. However, instead of growing full trees on subsets of the dataset, trees are grown to a predefined maximum depth using all of the data points. The path length of each data point ($q$) to its leaf node is then estimated using Eq.   \ref{eq:gen2out_path_length}, where $h_0$ is the path length up to the final node that the data point $q$ falls in, $l_{busy}$ is the number of points in that node, and $H(l_{busy})$ the estimated depth of an \textit{AtomTree} grown using $l_{busy}$ points.

\begin{equation}
    \label{eq:gen2out_path_length}
    h(q) = h_0 + H(l_{busy})
\end{equation}

The authors demonstrate that a linear relationship exists between the depth of the \textit{AtomTree} and the logarithm of the count of data points used to construct the tree, regardless of the distribution of the data. Based on this observation, a number of \textit{AtomTrees} are grown using several subsets of the data set to fit a linear function $H$ that maps the logarithm of the count of points to the depth of a fully grown \textit{AtomTree}. The anomaly score assigned to a point $q$ is then computed using Eq. \ref{eq:gen2out_anomaly_score}, where $n$ is the number of points in the considered data set and $E[h(q)]$ is the average path length of point $q$ in the forest.

\begin{equation} \label{eq:gen2out_anomaly_score}
    s(q, n) = 2^{-\frac{E[h(q)]}{H(n)}}
\end{equation}

\subsubsection{Sliding Histogram}
Our proposed approach specifically addresses anomalous groups defined in this study, i.e., dense groups of followers created in a tight time range. This is achieved by finding groups of followers that have a local distribution in the follower map that is significantly different from the overall distribution of the followers of the same user. The steps of this method are described as follows:
\begin{itemize}
    \item A window with a predefined width ($b$) is slid along the rank axis of the follower map. The window stretches on the timestamp axis between the lower and upper bounds of the follower timestamps at that position (Fig. \ref{fig:follower_map_and_sliding_hist}b).
    \item At each position, the window is divided into a predefined number of bins ($N_{bins}$) and the number of followers in each bin is computed (Fig. \ref{fig:follower_map_and_sliding_hist}c). These histograms are shown as line plots in Fig. \ref{fig:follower_map_and_sliding_hist}d.
    \item At each bin position, the median and inter-quartile range (IQR) of all histograms are computed.
    \item An anomaly score is assigned to each histogram bin using Eq. \ref{eq:bin_score}. Thus, each bin of followers is assigned a score that is the number of IQRs between the follower count in that bin and the median of follower counts in all bins at the same position.
    \begin{equation}
        A_{ij} = \frac{H_{ij} - M_{j} + 1}{IQR_j + 1}
    \label{eq:bin_score}
    \end{equation}
    
    Where $H_{ij}$ is the count of followers in the bin $j$ of the window $i$, and $M_{j}$ and $IQR_j$ are the median and IQR of follower counts in the bin $j$ across all windows, respectively.

    \item Since we are using a sliding window, each follower appears in more than one window. Thus, an anomaly score can be assigned to each individual follower $f$ using a weighted average of all bin scores $A_{ij}$ that include the follower $f$. The weight $\lambda_{fi}$ (Eq. \ref{eq:weight}) takes its maximum value when the follower $f$ is in the center of the bin and its minimum value when the follower $f$ is at the edge of the bin. The anomaly score is then computed using Eq. \ref{eq:anomaly_score}


    \begin{equation}
        \label{eq:weight}
        \lambda_{fi} = 1_{f \in W_i} \left( \frac{\frac{b}{2} - |R_f - C_i| + 1}{\sum_{j}\frac{b}{2} - |R_f - C_j| + 1} \right)
    \end{equation}

Where $b$ is the width of the sliding window, $R_f$ is the rank of the follower $f$, and $C$ is the center of the sliding window.

    \begin{equation}
        \label{eq:anomaly_score}
        score_f = \sum_j^{N_{bins}} \sum_i^{N_{windows}} \lambda_{fi} A_{ij} 1_{f \in W_i} 1_{f \in B_{ij}}
    \end{equation}
\end{itemize}

\FloatBarrier
\section{Results}

\subsection{Dribbble data results}
Table \ref{tab:results} shows the performance metrics of the anomaly detection methods using three different window sizes, averaged across all the synthetic Dribbble datasets. We evaluate the methods using the area under the ROC curve (AUC), average precision (AP), and precision when first 50 results considered (precision@50). ECOD performs best among the feature-based models. However, our suggested method clearly outperforms the feature-based methods in this task, especially when looking at the precision measures.

\begin{table}[H]
    \centering
    \begin{tabular}{lllll}
    \hline
    \multicolumn{1}{c}{Window} & \multicolumn{1}{c}{Method} &          \multicolumn{1}{c}{AUC} & \multicolumn{1}{c}{AP} & \multicolumn{1}{c}{P@50} \\
    \hline
    W51 & ECOD &  0.71 (0.22) &  0.31 (0.13) &  0.26 (0.21) \\
         & Gen2Out &  0.62 (0.31) &  0.26 (0.10) &  0.15 (0.18) \\
         & IsolationForest &  0.61 (0.30) &  0.24 (0.11) &  0.09 (0.17) \\
         & LocalOutlierFactor &  0.54 (0.20) &  0.28 (0.18) &  0.49 (0.31) \\
         & SlidingHistogram &  \textbf{0.86 (0.15)} &  \textbf{0.69 (0.23)} &  \textbf{0.72 (0.39)} \\
    \hline
    W101 & ECOD &  0.70 (0.21) &  0.29 (0.13) &  0.21 (0.19) \\
         & Gen2Out &  0.63 (0.30) &  0.25 (0.11) &  0.12 (0.18) \\
         & IsolationForest &  0.62 (0.28) &  0.23 (0.11) &  0.07 (0.16) \\
         & LocalOutlierFactor &  0.51 (0.19) &  0.25 (0.18) &  0.46 (0.37) \\
         & SlidingHistogram &  \textbf{0.87 (0.15)} &  \textbf{0.71 (0.23)} &  \textbf{0.72 (0.39)} \\
    \hline
    W201 & ECOD &  0.66 (0.21) &  0.26 (0.13) &  0.16 (0.17) \\
         & Gen2Out &  0.59 (0.27) &  0.22 (0.12) &  0.05 (0.11) \\
         & IsolationForest &  0.58 (0.26) &  0.20 (0.12) &  0.02 (0.09) \\
         & LocalOutlierFactor &  0.45 (0.17) &  0.21 (0.16) &  0.37 (0.38) \\
         & SlidingHistogram &  \textbf{0.87 (0.15)} &  \textbf{0.69 (0.24)} &  \textbf{0.70 (0.40)} \\
    \hline
\end{tabular}
    \caption{Area under ROC curve, average precision, and precision at 50 mean (std) values for all methods using different window sizes.}
    \label{tab:results}
\end{table}

\subsection{Twitter data results}
We apply the SH method on 1,318 accounts comprising Turkish politicians and media outlets from the \#Secim2023 dataset \cite{najafi2022secim2023} to explore the anomalous following patterns that this method can uncover. We use a window size of 200 since the results for window sizes 200 and 100 are similar, and since the politicians have a significantly larger number of followers than the Dribbble users. We divide our analysis of the Twitter dataset into three parts: (i) Retrieving user accounts that have anomalous followers (ii) Identifying individual anomalous follower accounts (iii) Exploring the coordinated behavior of the detected anomalous followers.

\subsubsection{Retrieving users with anomalous followers}
\begin{figure}[!htbp]
    \centering
    \includegraphics[width=\textwidth]{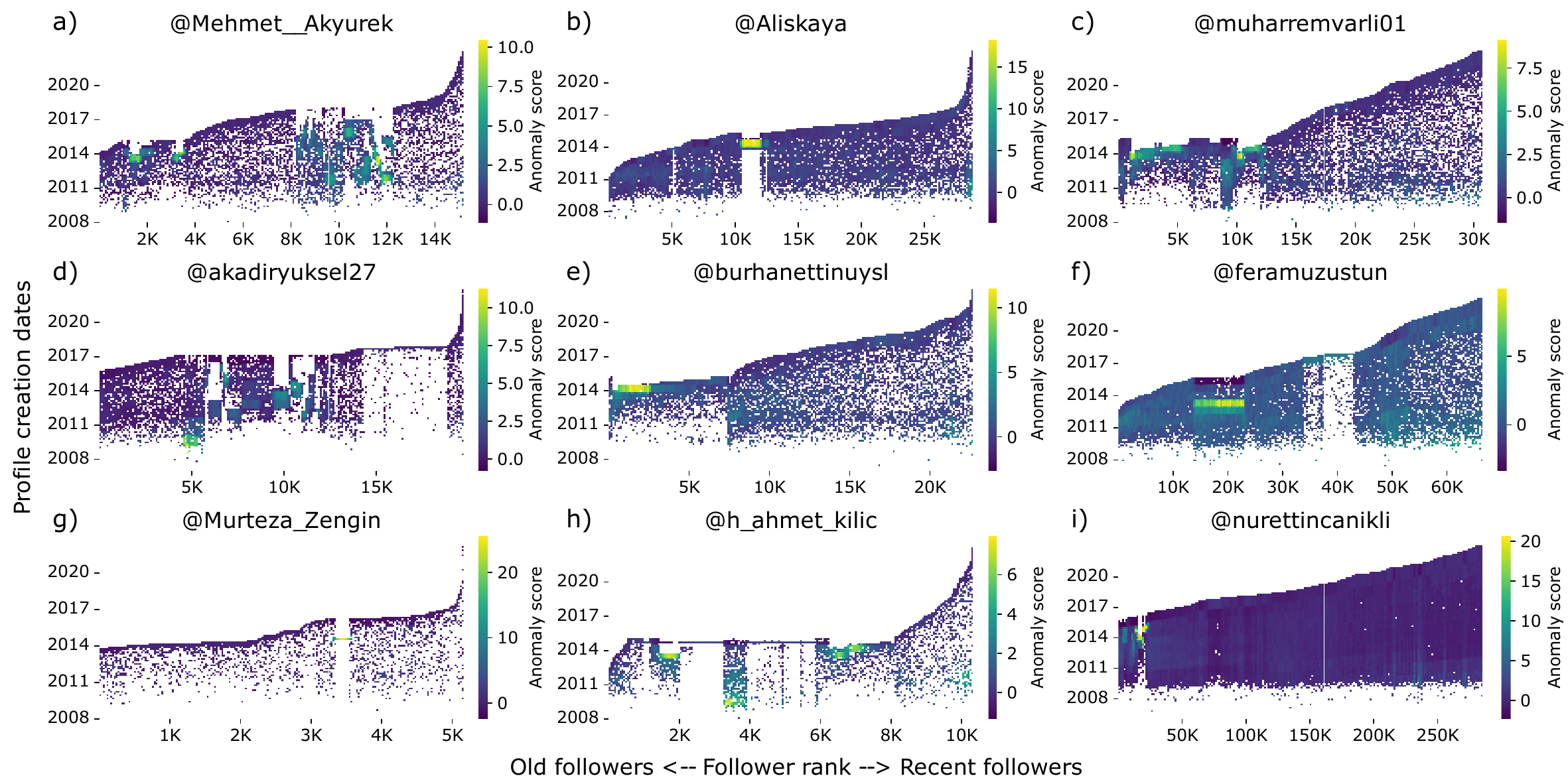}
    \caption{\textbf{Retrieving users with anomalous followers} Follower maps of the 9 Twitter accounts with the highest average anomaly score across all of their followers. The colors represent the average anomaly scores of all followers that fall in each bin (cell) of the heat map.}
    \label{fig:top_avg_anomaly_scores}
\end{figure}

In order to detect the users that have anomalous following patterns among their followers, we first look at the 9 Twitter accounts with the highest average anomaly score across all their followers (Fig. \ref{fig:top_avg_anomaly_scores}). We show the follower maps as heat maps instead of scatter plots since these users have high numbers of followers. Irregular following patterns can be observed in all of the follower maps of these users. Since the average anomaly score across all followers is generally lower for popular accounts, we can alternatively look at the average anomaly score of the highest N anomaly scores of a user's followers. Fig. \ref{fig:anomaly_and_bot_scores} shows the deviant followers of two popular Twitter accounts from our dataset. More examples can be seen in Appendix \ref{A:anomaly_samples}.

\subsubsection{Identifying individual anomalous accounts}
\begin{figure}[!htbp]
    \centering
    \includegraphics[width=\textwidth]{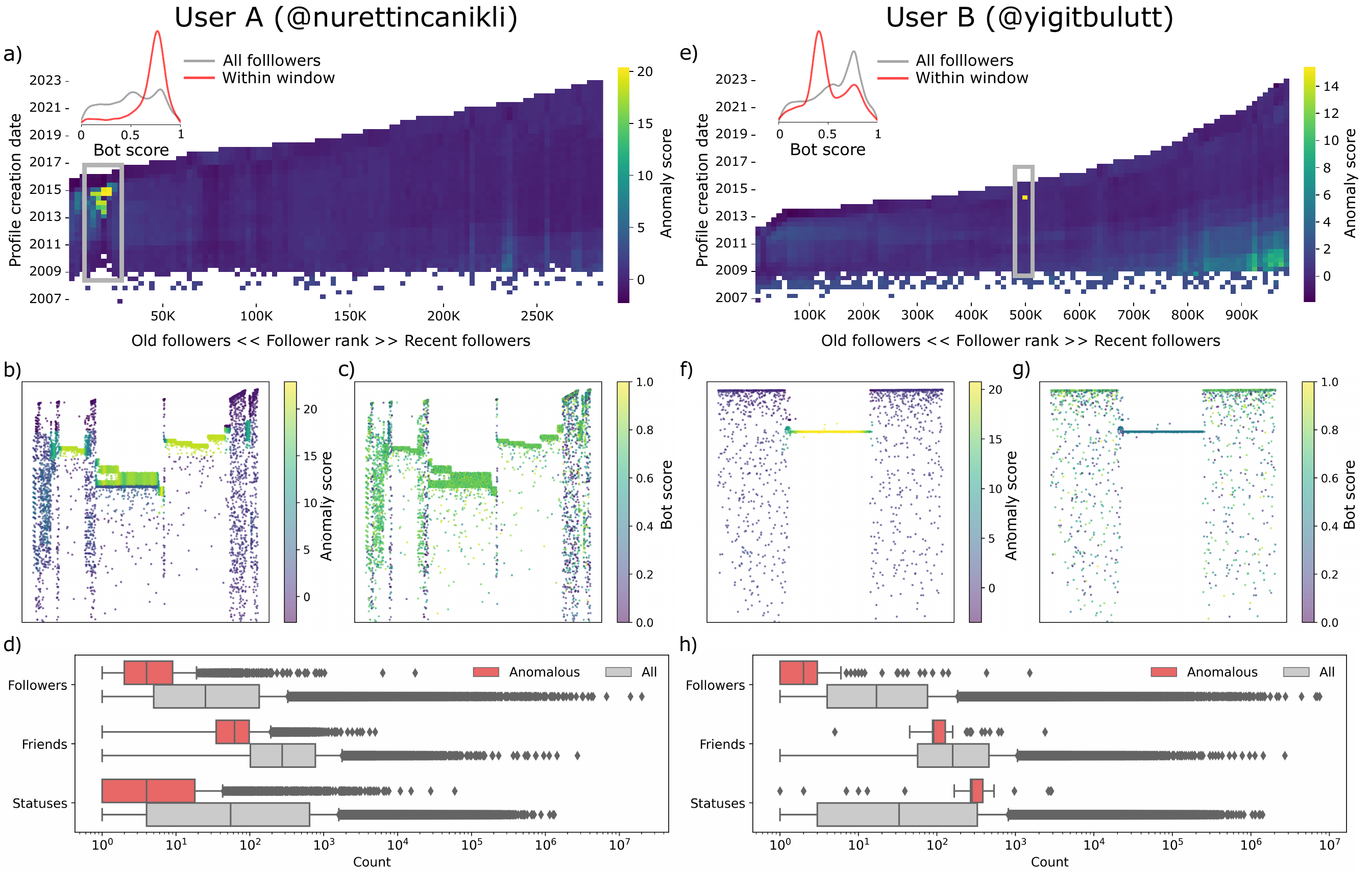}
    \caption{\textbf{Detailed analysis of anomalous followers} User A: Anomalous followers have high bot scores. User B: Anomalous followers have low bot scores. Anomalous regions are zoomed in for User A (b,c) and User B (f,g). Profile statistics for regular and all followers are also compared for these users in subplots (d) and (e).}
    \label{fig:anomaly_and_bot_scores}
\end{figure}

Next, we look at the individual accounts that constitute the groups of anomalous followers. First, we look at these accounts' bot scores as computed by BotometerLite \cite{yang2020scalable}. The BotometerLite only uses features that can be extracted from the account information, making it applicable to our dataset. 
We refer to the scores computed by the BotometerLite as \textit{bot scores}. Fig. \ref{fig:anomaly_and_bot_scores} shows two cases, (A) anomalous followers having high bot scores (B) anomalous followers having low bot scores. To validate that the anomalous followers in the second case are indeed suspicious accounts, we manually observe a sample of these accounts. Appendix \ref{A:anomalous_follower_profiles} shows samples of Twitter profiles of irregular followers of three accounts from our datasets, including the two accounts shown in Fig. \ref{fig:anomaly_and_bot_scores}. We observe that many of these accounts share the same tweets and share many of their friends. Additionally, the usernames of these accounts are in many cases meaningless combinations of letters. Fig. \ref{fig:anomaly_and_bot_scores}d and Fig. \ref{fig:anomaly_and_bot_scores}h show the distribution of the friend, follower, and status counts of the anomalous followers compared to that of all the followers of the same account. In both cases A and B, the anomalous accounts tend to have a lower number of followers. In case A, the anomalous followers have a low number of shared posts, indicating that they are mainly aimed at increasing the follower counts. On the other hand, the anomalous followers in case B share a lot of posts, indicating that they are used to spread information. These results show that our approach can capture bots that act in coordination, even though their bot scores as computed by other methods may not necessarily be high.

\subsubsection{Exploring anomalous follower group behavior}
\begin{figure}[!htbp]
    \centering
    \includegraphics[width=\textwidth]{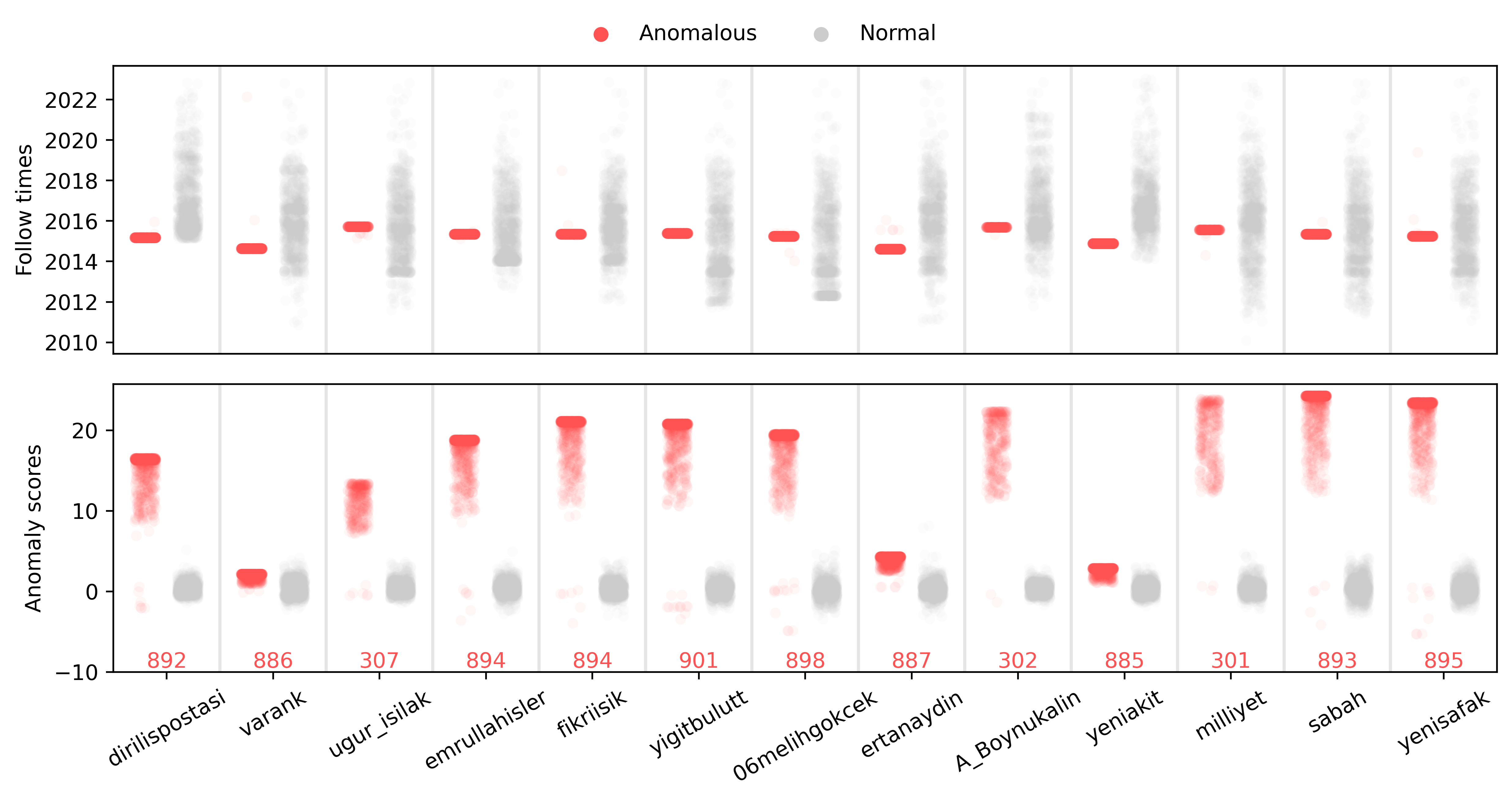}
    \caption{\textbf{Coordinated behavior of anomalous followers} Follow times (top) and anomaly scores (bottom) of the shared anomalous followers (red) and the shared non-anomalous followers (gray) across 13 users that are followed by the same batch of anomalous followers shown in Fig.\ref{fig:anomaly_and_bot_scores}(f).}
    \label{fig:follower_analysis1}
\end{figure}

We explore the following patterns of the detected groups of anomalous followers and study when they follow other users in our Twitter dataset. Are they always showing suspicious following patterns for other politicians, or is it specific to the particular user that we made the observation? 
Firstly, we look for accounts in our dataset that are followed by at least 30\% of the suspicious followers of users A and B (Fig.\ref{fig:anomaly_and_bot_scores}). We find 0 accounts followed by the anomalous followers of user A and 12 accounts followed by the anomalous followers of user B. Since the anomalous accounts following user A do not follow any other users from our dataset, we resume our analysis for user B only. We estimate the dates that the anomalous followers followed each of the 13 Twitter accounts using the method suggested in \cite{meeder2011we}. Appendix presents results for evaluation of the follow time estimation method on the Dribbble dataset, which provides ground truth values for follow times. Fig.\ref{fig:follower_analysis1}(a) and Fig.\ref{fig:follower_analysis1}(b) show the following times and anomaly scores, respectively, of the anomalous followers (red) and the followers shared across the 13 users (gray) for comparison. The anomalous followers follow each user almost simultaneously, which demonstrates that they are automated accounts that work in coordination. Furthermore, the anomalous followers followed all of the 13 users between the years 2014 and 2016. Finally, our approach correctly assigned high anomaly scores to the anomalous followers in most cases (Fig.\ref{fig:follower_analysis1}(b)).

\begin{figure}[!htbp]
    \centering
    \includegraphics[width=\textwidth]{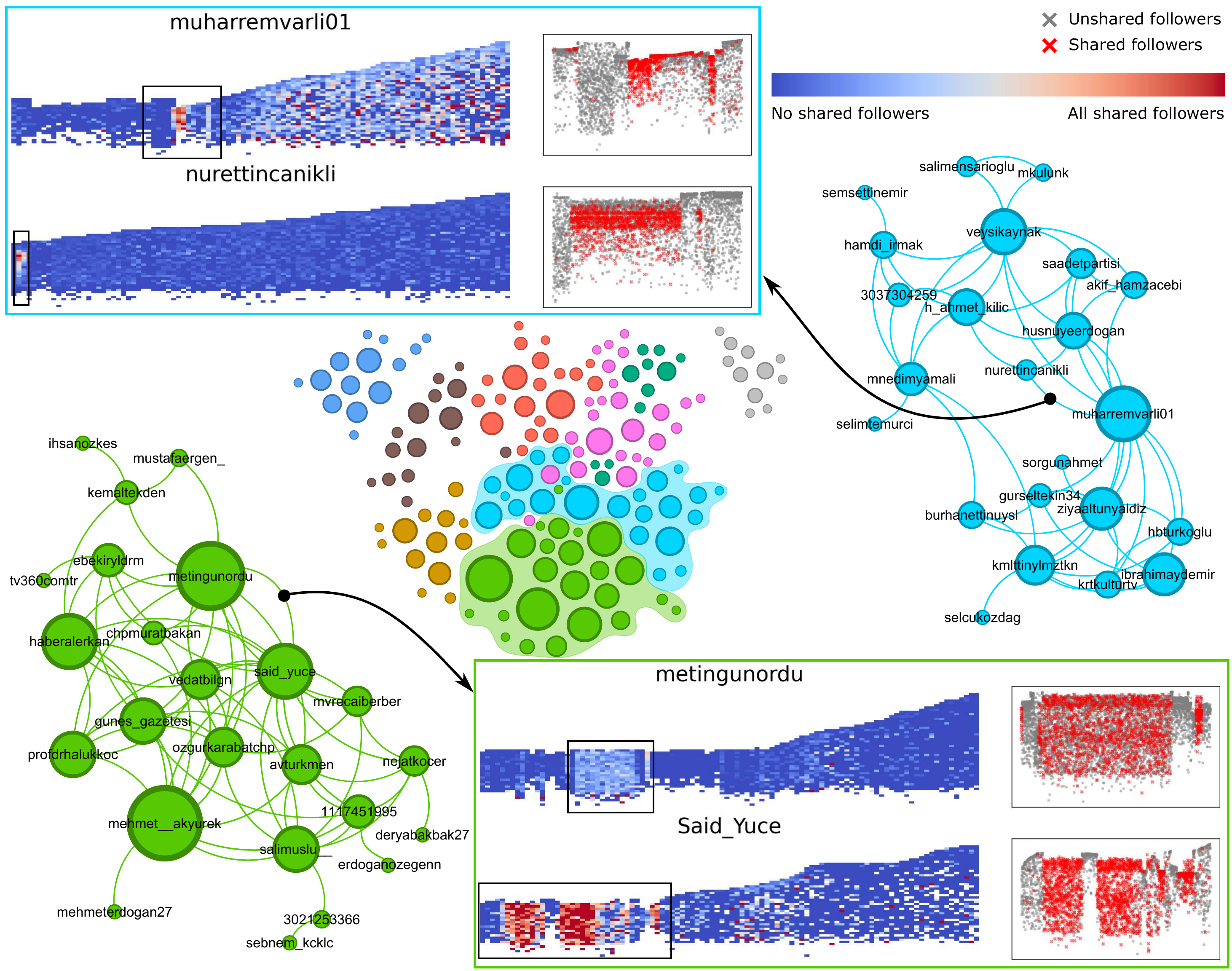}
    \caption{\textbf{Similarity network based on the shared anomalous followers.} The full network is shown in the middle of the figure, where nodes are colored based on communities and sized based on their degrees. The two communities with the highest pairwise average anomaly scores are highlighted and shown in detail along with the follower maps of one edge in each community.}
    \label{fig:network_analysis}
\end{figure}

We expand the analysis of the group behavior of anomalous followers to uncover other groups of accounts that share the same suspicious followers. For this purpose, we create a similarity network based on the shared anomalous followers. The similarity between each pair of users is the cosine similarity between the two anomaly score vectors of the followers shared across the pair of users. Since our method assigns anomaly scores based on the follower map, a follower that follows users U1 and U2 will have two different anomaly scores computed for U1 and U2. Thus, a pair of users that share followers who were assigned high anomaly scores in both follower maps will have a high similarity. Fig.\ref{fig:network_analysis} shows the two communities with the highest pairwise average anomaly scores across all edges in the community. For each community, we show the follower maps of a user pair corresponding to one of the edges in the community. The follower maps are colored by the ratio of shared followers between the pair of users in each bin. This allows us to capture concentrations of shared followers in both users' maps, which appear as reddish regions in the follower map. We observe that the concentrated regions of shared followers exhibit anomalous following patterns in both follower maps. This finding supports our hypothesis that anomalous followers work in coordination. More details about this network analysis and other samples of anomalous follower groups appearing in different users' follower maps are presented in Appendix \ref{A:shared_anomalous_followers}.

\FloatBarrier
\section{Discussion}

Our analysis of the followers of 1,318 Twitter accounts supports the earlier findings regarding the existence of anomalous following patterns on Twitter \footnote{https://www.nytimes.com/interactive/2018/01/27/technology/social-media-bots.html} \cite{varol2020journalists}. 
Especially exogenous events like political campaigns and elections can change composition of followers and Twitter accounts gain new followers \cite{varol2023follows}.
Our manual observation of samples of the accounts that engage in anomalous following patterns shows that these accounts are indeed fake accounts. Moreover, we observe that batches of anomalous followers tend to follow Twitter accounts almost simultaneously, suggesting that they are automated accounts managed from one place. The fact that many of the detected anomalous followers in this study are still active accounts indicates that Twitter has not yet identified these accounts as malicious. Although these are clearly fake and automated accounts, we cannot make any conclusions about their intentions. We hypothesize three possible scenarios: (i) The user purchased these followers to gain popularity (ii) The anomalous accounts followed the user to gain credibility or fit in a specific persona (iii) The user was targeted by the anomalous followers to serve a propaganda for the user's opponents.

The main limitation of our approach is that it tend to compute anomaly scores that lead higher false negatives when the ratio of anomalous followers is high. SH assigns high anomaly scores to followers that deviate from the main follower distribution, as defined by the set of histograms. Therefore, the method will not assign the right anomaly scores when the account's followers are dominated by anomalous followers. Fig.\ref{fig:high_anomaly_ratio} shows some cases with high ratios of anomalous followers, where the scores are wrongly assigned.

The recent changes in the API policy has rendered Twitter data less accessible. However, our method is applicable to any other social media platform that provides an ordered list of followers and their creation dates. Furthermore, detecting this type of behavior allows for understanding coordinated activities and misinformation-spreading campaigns that may have happened earlier on Twitter.

Detecting coordinated misinformation campaigns on social media platforms has become crucial in the recent years \cite{zhang2023capturing}. In this paper, we present a method to detect a previously unaddressed type of anomalous followers on social media platforms. We demonstrate that the detected anomalous followers act in coordination and in many cases exhibit similar anomalous behavior across more than one account. Using this approach, further analyses can be applied to uncover coordinated misinformation activities on social media platforms.

\FloatBarrier
\section{Acknowledgements}
We thank Rossano Schifanella for providing access to Dribbble dataset. We also thank Baris Temel for his earlier work on the topic.
This work is partly supported by TUBITAK projects 121C220 and 222N311. 


\FloatBarrier
\bibliographystyle{unsrt}
\bibliography{main}

\begin{thebibliography}{10}

\bibitem{ferrara2016rise}
Emilio Ferrara, Onur Varol, Clayton Davis, Filippo Menczer, and Alessandro
  Flammini.
\newblock The rise of social bots.
\newblock {\em Communications of the ACM}, 59(7):96--104, 2016.

\bibitem{cresci2020decade}
Stefano Cresci.
\newblock A decade of social bot detection.
\newblock {\em Communications of the ACM}, 63(10):72--83, 2020.

\bibitem{alkulaib2022twitter}
Lulwah Alkulaib, Lei Zhang, Yanshen Sun, and Chang-Tien Lu.
\newblock Twitter bot identification: An anomaly detection approach.
\newblock In {\em 2022 IEEE International Conference on Big Data (Big Data)},
  pages 3577--3585. IEEE, 2022.

\bibitem{yang2019arming}
Kai-Cheng Yang, Onur Varol, Clayton~A Davis, Emilio Ferrara, Alessandro
  Flammini, and Filippo Menczer.
\newblock Arming the public with artificial intelligence to counter social
  bots.
\newblock {\em Human Behavior and Emerging Technologies}, 1(1):48--61, 2019.

\bibitem{bruno2022brexit}
Matteo Bruno, Renaud Lambiotte, and Fabio Saracco.
\newblock Brexit and bots: characterizing the behaviour of automated accounts
  on twitter during the uk election.
\newblock {\em EPJ Data Science}, 11(1):17, 2022.

\bibitem{varol2020journalists}
Onur Varol and Ismail Uluturk.
\newblock Journalists on twitter: self-branding, audiences, and involvement of
  bots.
\newblock {\em Journal of Computational Social Science}, 3(1):83--101, 2020.

\bibitem{shao2018spread}
Chengcheng Shao, Giovanni~Luca Ciampaglia, Onur Varol, Kai-Cheng Yang,
  Alessandro Flammini, and Filippo Menczer.
\newblock The spread of low-credibility content by social bots.
\newblock {\em Nature communications}, 9(1):1--9, 2018.

\bibitem{himelein2021bots}
McKenzie Himelein-Wachowiak, Salvatore Giorgi, Amanda Devoto, Muhammad Rahman,
  Lyle Ungar, H~Andrew Schwartz, David~H Epstein, Lorenzo Leggio, and Brenda
  Curtis.
\newblock Bots and misinformation spread on social media: Implications for
  covid-19.
\newblock {\em Journal of medical Internet research}, 23(5):e26933, 2021.

\bibitem{varol2017online}
Onur Varol, Emilio Ferrara, Clayton Davis, Filippo Menczer, and Alessandro
  Flammini.
\newblock Online human-bot interactions: Detection, estimation, and
  characterization.
\newblock In {\em Proceedings of the international AAAI conference on web and
  social media}, volume~11, pages 280--289, 2017.

\bibitem{varol2023should}
Onur Varol.
\newblock Should we agree to disagree about twitter’s bot problem?
\newblock {\em Online Social Networks and Media}, 37:100263, 2023.

\bibitem{sayyadiharikandeh2020detection}
Mohsen Sayyadiharikandeh, Onur Varol, Kai-Cheng Yang, Alessandro Flammini, and
  Filippo Menczer.
\newblock Detection of novel social bots by ensembles of specialized
  classifiers.
\newblock In {\em Proceedings of the 29th ACM international conference on
  information \& knowledge management}, pages 2725--2732, 2020.

\bibitem{liu2023botmoe}
Yuhan Liu, Zhaoxuan Tan, Heng Wang, Shangbin Feng, Qinghua Zheng, and Minnan
  Luo.
\newblock Botmoe: Twitter bot detection with community-aware mixtures of
  modal-specific experts.
\newblock {\em arXiv preprint arXiv:2304.06280}, 2023.

\bibitem{ding2023find}
Jianwei Ding and Zhouguo Chen.
\newblock How to find social robots exactly?
\newblock In {\em Proceedings of the 2023 6th International Conference on
  Software Engineering and Information Management}, pages 12--18, 2023.

\bibitem{echeverri2018lobo}
Juan Echeverr{\"\i}{\pounds}!`~a, Emiliano De~Cristofaro, Nicolas Kourtellis,
  Ilias Leontiadis, Gianluca Stringhini, and Shi Zhou.
\newblock Lobo: Evaluation of generalization deficiencies in twitter bot
  classifiers.
\newblock In {\em Proceedings of the 34th annual computer security applications
  conference}, pages 137--146, 2018.

\bibitem{jia2017random}
Jinyuan Jia, Binghui Wang, and Neil~Zhenqiang Gong.
\newblock Random walk based fake account detection in online social networks.
\newblock In {\em 2017 47th annual IEEE/IFIP international conference on
  dependable systems and networks (DSN)}, pages 273--284. IEEE, 2017.

\bibitem{mendoza2020bots}
Marcelo Mendoza, Maurizio Tesconi, and Stefano Cresci.
\newblock Bots in social and interaction networks: detection and impact
  estimation.
\newblock {\em ACM Transactions on Information Systems (TOIS)}, 39(1):1--32,
  2020.

\bibitem{mazza2019rtbust}
Michele Mazza, Stefano Cresci, Marco Avvenuti, Walter Quattrociocchi, and
  Maurizio Tesconi.
\newblock Rtbust: Exploiting temporal patterns for botnet detection on twitter.
\newblock In {\em Proceedings of the 10th ACM conference on web science}, pages
  183--192, 2019.

\bibitem{mannocci2022mulbot}
Lorenzo Mannocci, Stefano Cresci, Anna Monreale, Athina Vakali, and Maurizio
  Tesconi.
\newblock Mulbot: Unsupervised bot detection based on multivariate time series.
\newblock In {\em 2022 IEEE International Conference on Big Data (Big Data)},
  pages 1485--1494. IEEE, 2022.

\bibitem{meeder2011we}
Brendan Meeder, Brian Karrer, Amin Sayedi, R~Ravi, Christian Borgs, and
  Jennifer Chayes.
\newblock We know who you followed last summer: inferring social link creation
  times in twitter.
\newblock In {\em Proceedings of the 20th international conference on World
  wide web}, pages 517--526, 2011.

\bibitem{najafi2022secim2023}
Ali Najafi, Nihat Mugurtay, Ege Demirci, Serhat Demirkiran, Huseyin~Alper
  Karadeniz, and Onur Varol.
\newblock \#secim2023: First public dataset for studying turkish general
  election, 2022.

\bibitem{liu2008isolation}
Fei~Tony Liu, Kai~Ming Ting, and Zhi-Hua Zhou.
\newblock Isolation forest.
\newblock In {\em 2008 eighth ieee international conference on data mining},
  pages 413--422. IEEE, 2008.

\bibitem{breunig2000lof}
Markus~M Breunig, Hans-Peter Kriegel, Raymond~T Ng, and J{\"o}rg Sander.
\newblock Lof: identifying density-based local outliers.
\newblock In {\em Proceedings of the 2000 ACM SIGMOD international conference
  on Management of data}, pages 93--104, 2000.

\bibitem{li2022ecod}
Zheng Li, Yue Zhao, Xiyang Hu, Nicola Botta, Cezar Ionescu, and George Chen.
\newblock Ecod: Unsupervised outlier detection using empirical cumulative
  distribution functions.
\newblock {\em IEEE Transactions on Knowledge and Data Engineering}, 2022.

\bibitem{lee2021gen}
Meng-Chieh Lee, Shubhranshu Shekhar, Christos Faloutsos, T~Noah Hutson, and
  Leon Iasemidis.
\newblock Gen 2 out: Detecting and ranking generalized anomalies.
\newblock In {\em 2021 IEEE International Conference on Big Data (Big Data)},
  pages 801--811. IEEE, 2021.

\bibitem{yang2020scalable}
Kai-Cheng Yang, Onur Varol, Pik-Mai Hui, and Filippo Menczer.
\newblock Scalable and generalizable social bot detection through data
  selection.
\newblock In {\em Proceedings of the AAAI conference on artificial
  intelligence}, volume~34, pages 1096--1103, 2020.

\bibitem{varol2023follows}
Onur Varol.
\newblock Who follows turkish presidential candidates in 2023 elections?
\newblock In {\em 2023 31st Signal Processing and Communications Applications
  Conference (SIU)}, pages 1--4. IEEE, 2023.

\bibitem{zhang2023capturing}
Yizhou Zhang, Karishma Sharma, and Yan Liu.
\newblock Capturing cross-platform interaction for identifying coordinated
  accounts of misinformation campaigns.
\newblock In {\em European Conference on Information Retrieval}, pages
  694--702. Springer, 2023.

\bibitem{blondel2008fast}
Vincent~D Blondel, Jean-Loup Guillaume, Renaud Lambiotte, and Etienne Lefebvre.
\newblock Fast unfolding of communities in large networks.
\newblock {\em Journal of statistical mechanics: theory and experiment},
  2008(10):P10008, 2008.

\end{thebibliography}

\pagebreak

\setcounter{figure}{0}
\renewcommand{\thefigure}{SI-\arabic{figure}}
\setcounter{table}{0}
\renewcommand{\thetable}{SI-\arabic{table}}

\FloatBarrier
\appendix
\section{Appendices}

\subsection{Data details and preparation} \label{A:data_details}
\begin{figure}[H]
    \centering
    \includegraphics[width=\textwidth]{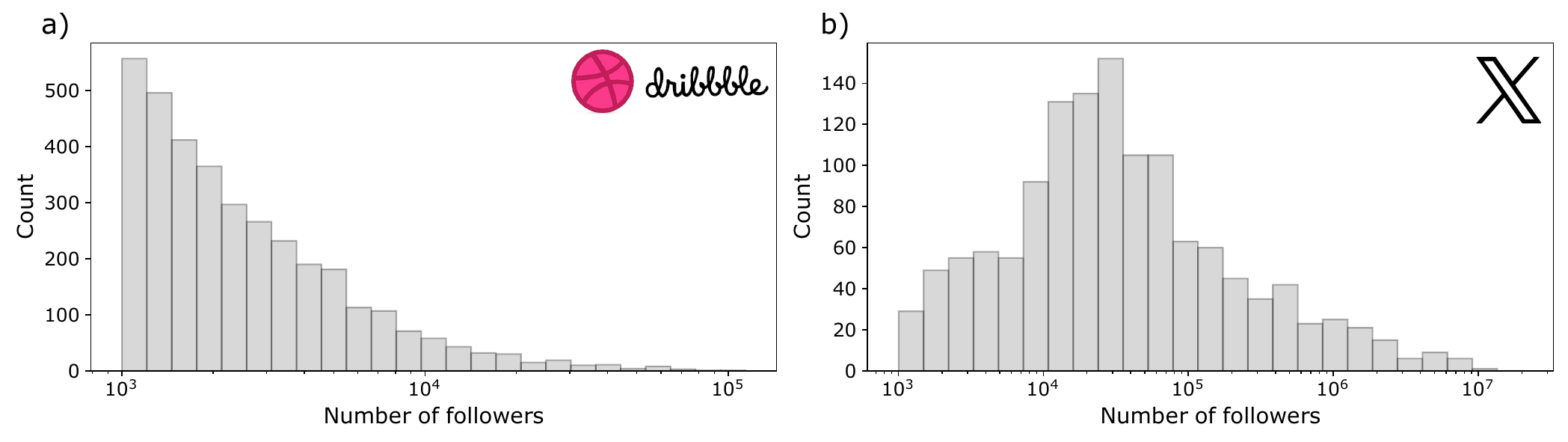}
    \caption{Follower count distributions in the Dribbble (a) and Twitter (b) datasets}
    \label{fig:follower_dist}
\end{figure}

Fig. \ref{fig:follower_dist} shows the follower distributions in each of the Dribbble and Twitter datasets used in this study. In both datasets, we only include users with at least 1,000 followers.

\begin{figure}[H]
    \centering
    \includegraphics[width=\textwidth]{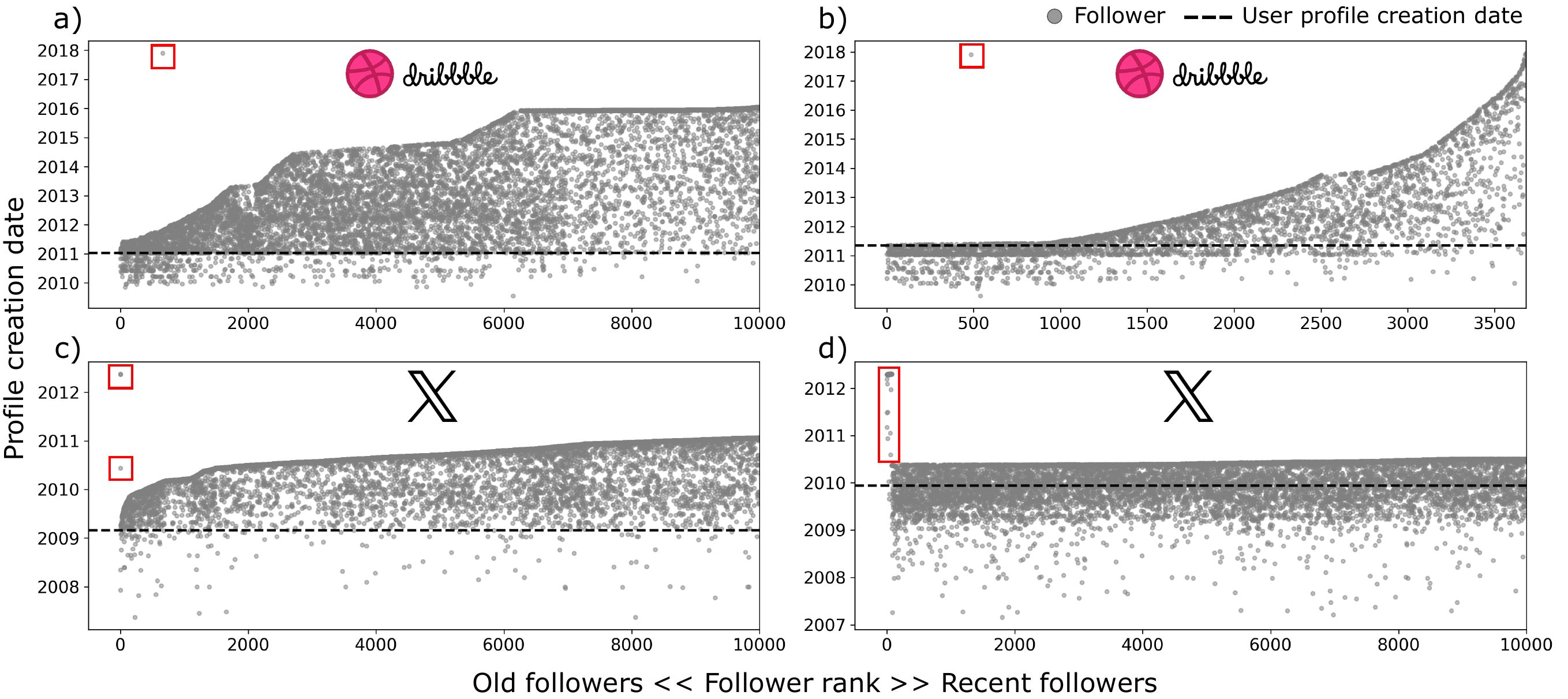}
    \caption{\textbf{Misplaced followers} in the Dribbble (a-b) and Twitter (c-d) datasets. The misplaced followers are marked by red boxes.}
    \label{fig:rank_problem}
\end{figure}

We noticed that some users in both datasets had followers that were clearly above the upper bound (Fig. \ref{fig:rank_problem}). The misplaced followers can either be users who were wrongly placed in the ordered follower list provided by the social media platform or their profile creation dates were inaccurate. We removed these followers since they affect the follow time estimation and anomaly detection algorithms. Removing these cases from the Dribbble dataset was simple since, given the ground truth follow times, we can look for large errors in the estimated follow times. As for the Twitter dataset, we first retrieved the users that have a sudden change in the upper bound that exceeds a threshold (90 users). Then, we manually examined the follower maps of the retrieved users and removed the misplaced followers. In total, there were 2 users in the Dribbble dataset and 20 users in the Twitter dataset that had misplaced followers. We noticed that all of the 20 user accounts from the Twitter dataset were created before 2012.

\subsection{Synthetic data generation} \label{A:synthetic_data_generation}
Type 1 followers consist of a batch of $N_1$ followers that were created in a range of time sampled from a normal distribution $\mathcal{N}(t_0, \sigma)$, where $t_0$ is randomly sampled from the range of timestamps where the follower batch is inserted. The rank in which type 1 synthetic followers are inserted is randomly sampled between the 10th and 90th follower rank percentiles.
Type 2 followers are created by replicating each of the last $N_{recent}$ followers that fall on the upper bound of the follower map $N_{replica}$ times, resulting in a total number of synthetic followers $N_2 = N_{recent} * N_{replica}$
The values used to create the different permutations of each synthetic type are presented in Table \ref{tab:synthetic_follower_perm}. In the permutations where both anomaly types exist, we include an equal number of anomalies from each type.

\begin{figure}[H]
    \centering
    \includegraphics[width=\textwidth]{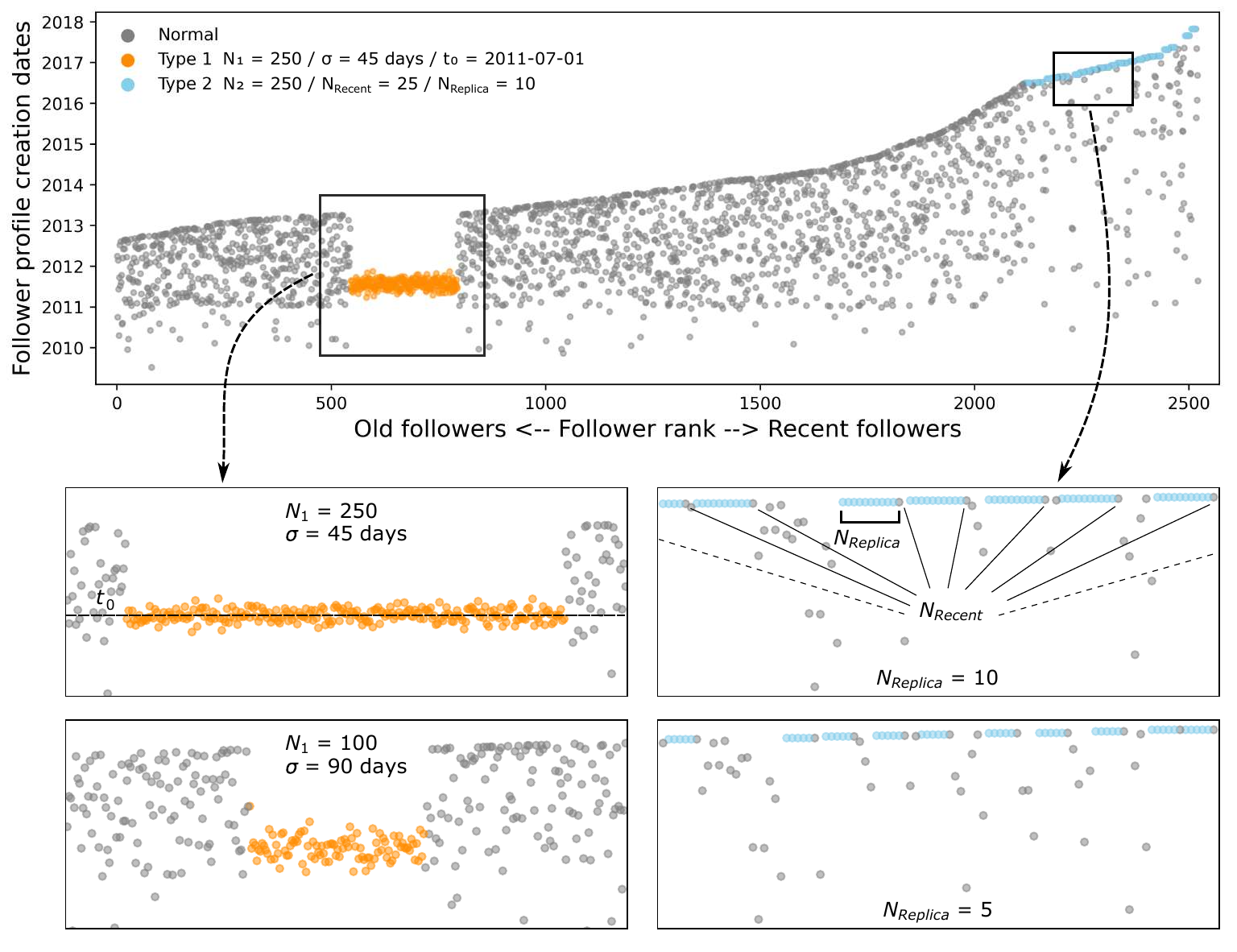}
    \caption{Illustration of the inserted synthetic anomalous followers.}
    \label{fig:synthetic_data}
\end{figure}

\begin{table}[H]
    \centering
    \begin{tabular}{|c|c|c|}
    \hline
       Type 1  &  $N_1$ & 50, 100, 250, 500, 1000 \\
               & $\sigma$ & 10, 45, 90 days \\
    \hline 
       Type 2  &  $N_2$ & 50, 100, 250, 500, 1000 \\
               & $N_{replica}$ & 5, 10 \\
               \hline
    \end{tabular}
    \caption{Parameter values for synthetic follower generation.}
    \label{tab:synthetic_follower_perm}
\end{table}

\subsection{Feature engineering} \label{A:feature_engineering}
\begin{figure}[H]
    \centering
    \includegraphics[width=\textwidth]{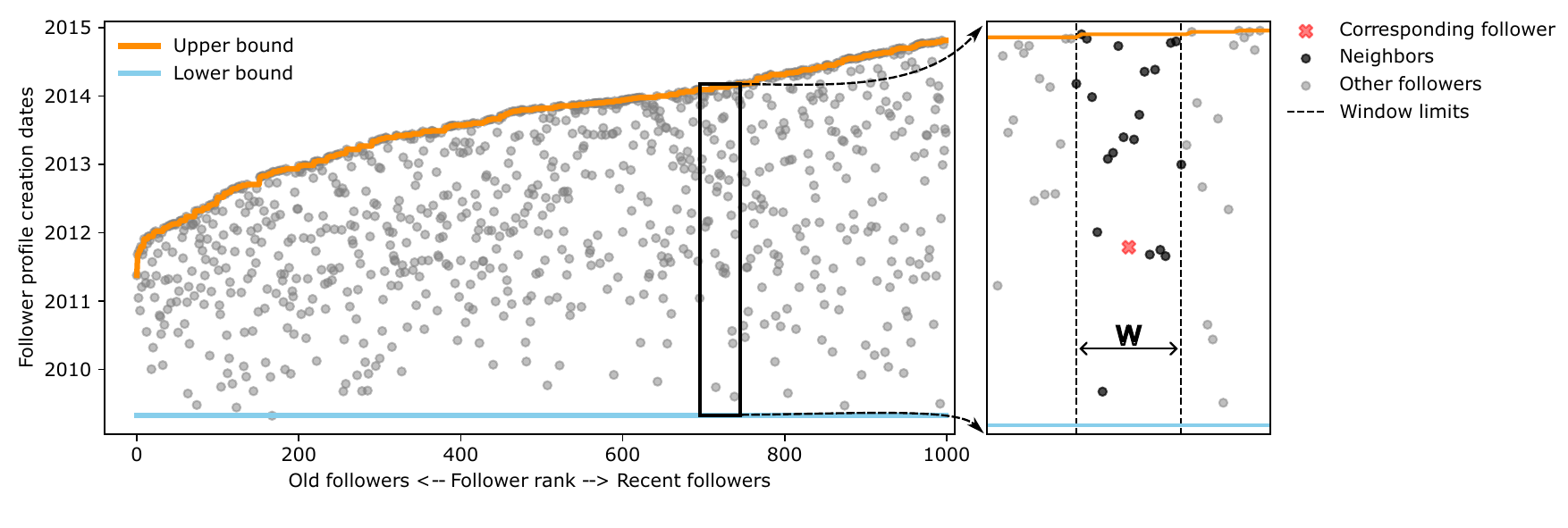}
    \caption{Profile creation date lower and upper bounds, and centered window around the corresponding follower for which features are computed}
    \label{fig:feature_eng}
\end{figure}

The features used to detect anomalous followers using anomaly detection algorithms are described in Table \ref{tab:feature_def}. Fig. \ref{fig:feature_eng} demonstrates the lower and upper bounds of a follower map, in addition to the window used to compute features that are based on the neighbors of a follower.

\begin{table}[H]
    \centering
    \begin{tabular}{p{0.4\linewidth}  p{0.55\linewidth}}
    \hline
    Feature & Description \\
    \hline
    Avg. neighbor creation date & The average profile creation date of the neighbors in a centered window of width W, weighted by the rank difference between each neighbor and the corresponding follower \\

     & \\
    
    Neighbor creation date range & Difference between the 90th and 10th percentiles of the creation dates of the neighbors in a centered window of width W  \\

     & \\
    
    Avg. distance to neighbors & Average distance to the neighbors in a centered window of width W, measured in terms of creation dates and weighted by the rank difference between each neighbor and the corresponding follower\\

     & \\
    
    Creation date boundary range & The difference between the lower and upper bounds of profile creation dates at the rank of the corresponding follower \\

     & \\
    
    Distance to upper bound & Difference between the profile creation date upper bound and the profile creation date of the corresponding follower \\

     & \\
    
    Relative rank & Rank of the corresponding follower divided by the total number of followers \\
    
    \hline
    \end{tabular}
    \caption{Unsupervised anomaly detection feature definitions}
    \label{tab:feature_def}
\end{table}

\subsection{Evaluation of the follow time estimation method}
We conduct an evaluation of the follow-time estimation method suggested by \cite{meeder2011we} by comparing the estimated follow time to the ground truth follow time provided by the Dribbble platform. As expected, the error drops as the number of followers increases.
\begin{figure}[H]
    \centering
    \includegraphics[width=\textwidth]{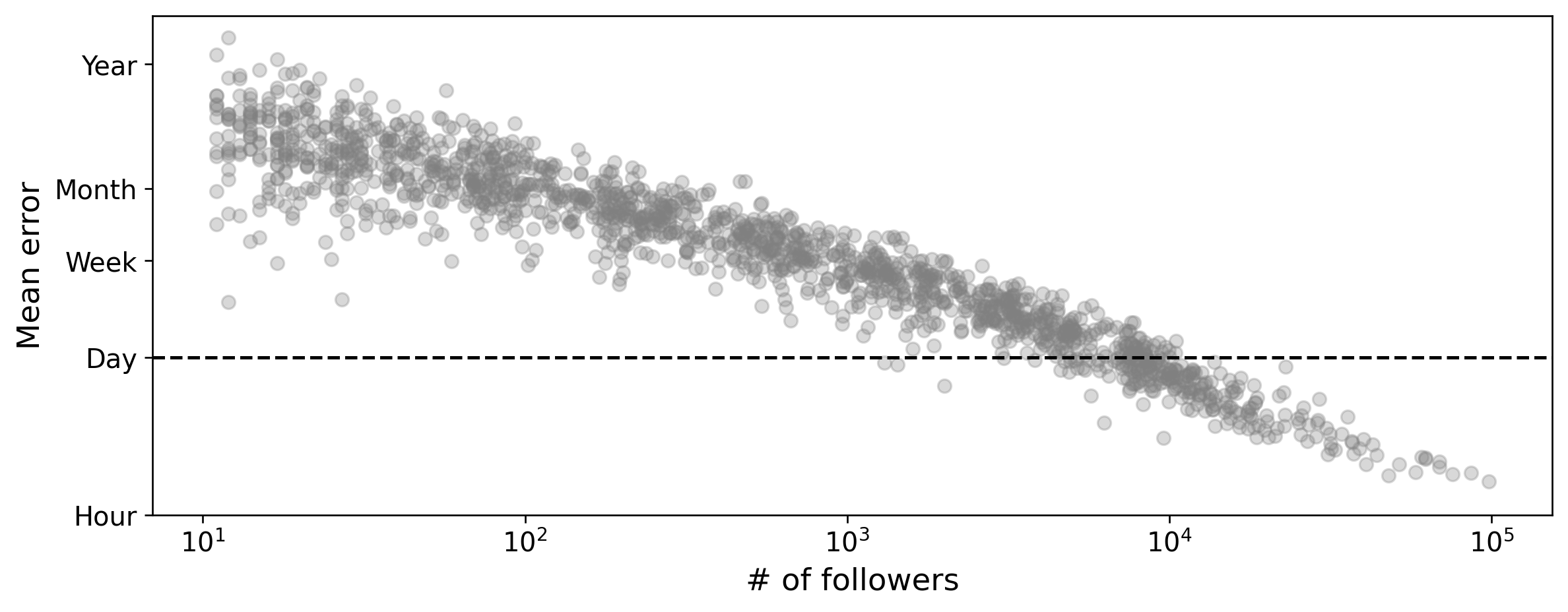}
    \caption{\textbf{Mean follow time estimation error} Each point represents the error between the estimated follow time and the ground truth averaged across all followers of one Dribbble user. The mean error is less than one day for users with more than 10,000 followers.}
    \label{fig:obvious_anomalies}
\end{figure}

\subsection{Samples of detected anomalous followers} \label{A:anomaly_samples}

\begin{figure}[H]
    \centering
    \includegraphics[width=\textwidth]{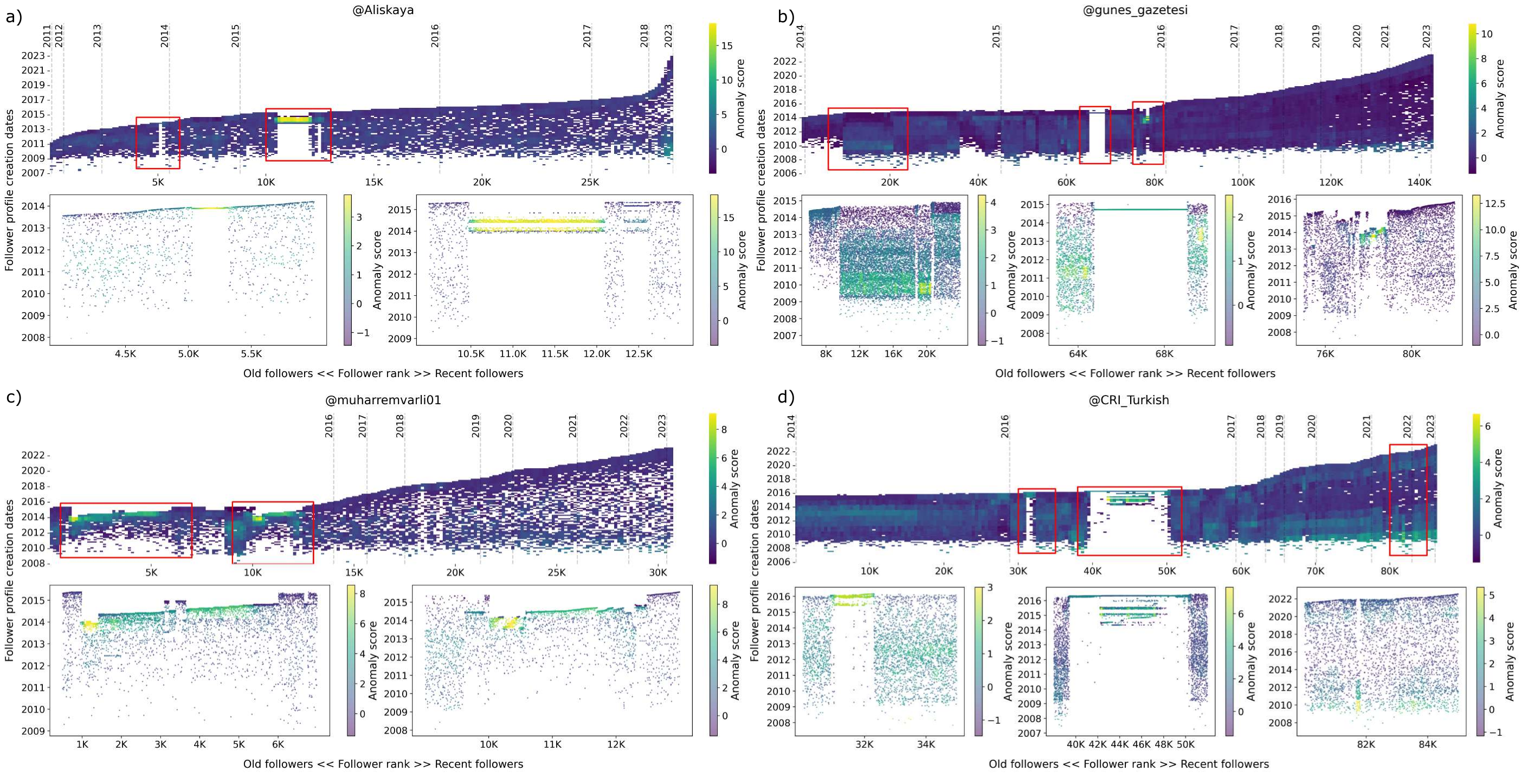}
    \caption{\textbf{Samples of distinct anomalous following patterns}}
    \label{fig:obvious_anomalies}
\end{figure}

\begin{figure}[H]
    \centering
    \includegraphics[width=\textwidth]{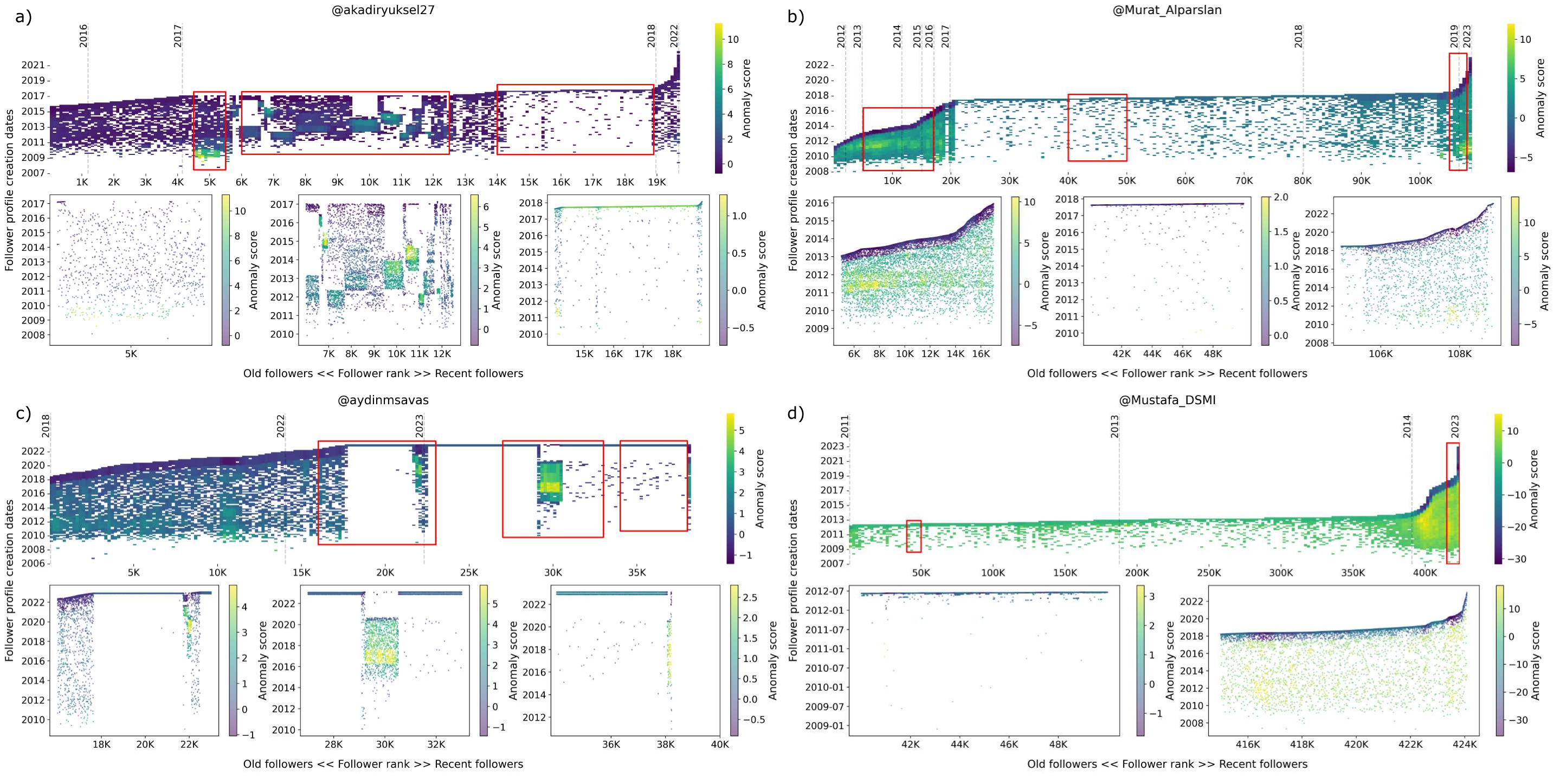}
    \caption{\textbf{Samples of follower maps dominated by anomalous following patterns} Anomaly scores are wrongly assigned due to the high ratio of anomalous followers}
    \label{fig:high_anomaly_ratio}
\end{figure}

\begin{figure}[H]
    \centering
    \includegraphics[width=\textwidth]{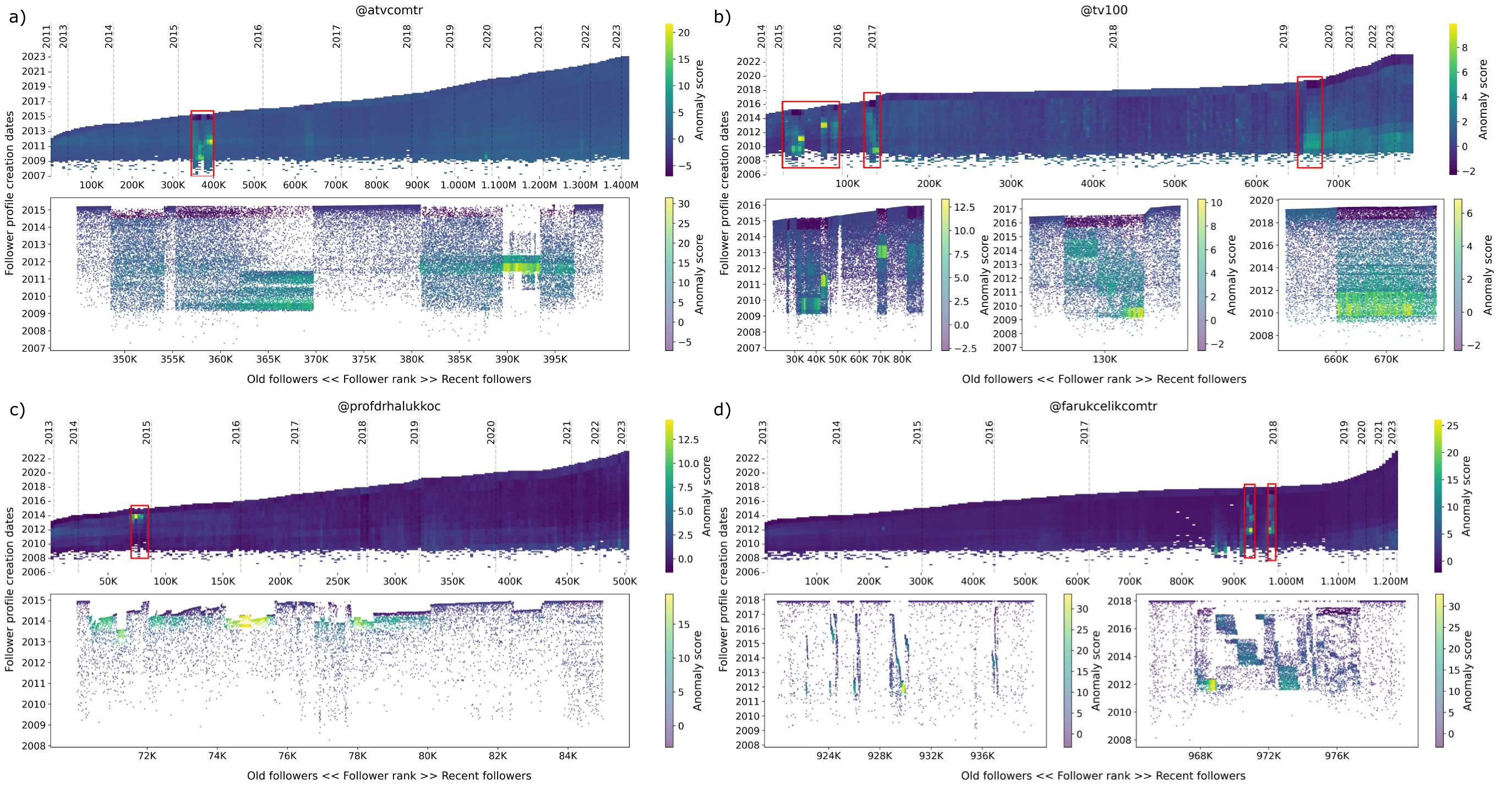}
    \caption{\textbf{Samples of anomalous following patterns in popular accounts}}
    \label{fig:high_anomaly_ratio}
\end{figure}

\subsection{Shared anomalous followers} \label{A:shared_anomalous_followers}
\begin{figure}[H]
    \centering
    \includegraphics[width=\textwidth]{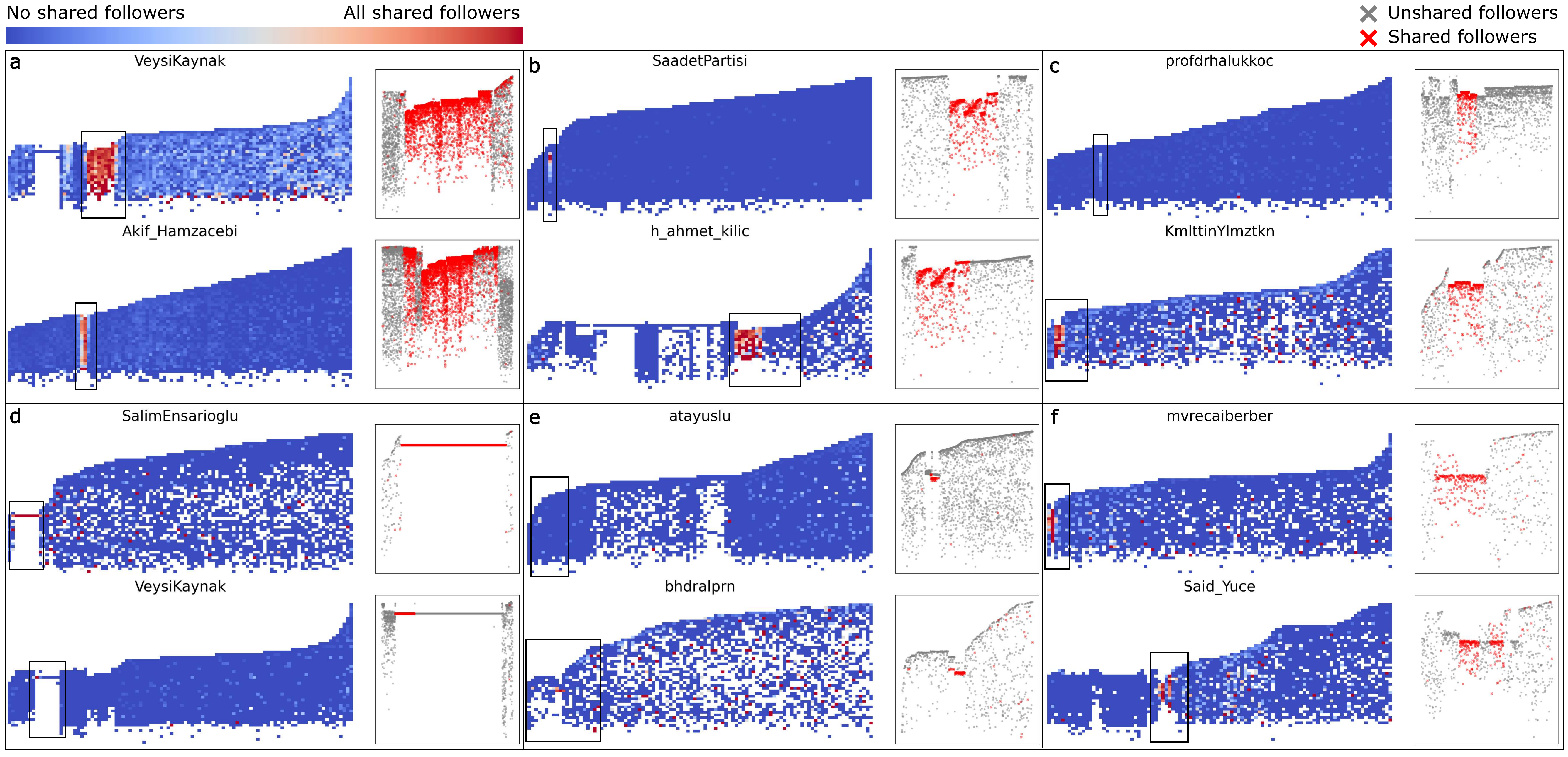}
    \caption{\textbf{Shared anomalous followers} Follower maps of the 6 user pairs corresponding to the highest similarity scores in our dataset.}
    \label{fig:top_similarity_scores}
\end{figure}

\begin{figure}[H]
    \centering
    \includegraphics[width=\textwidth]{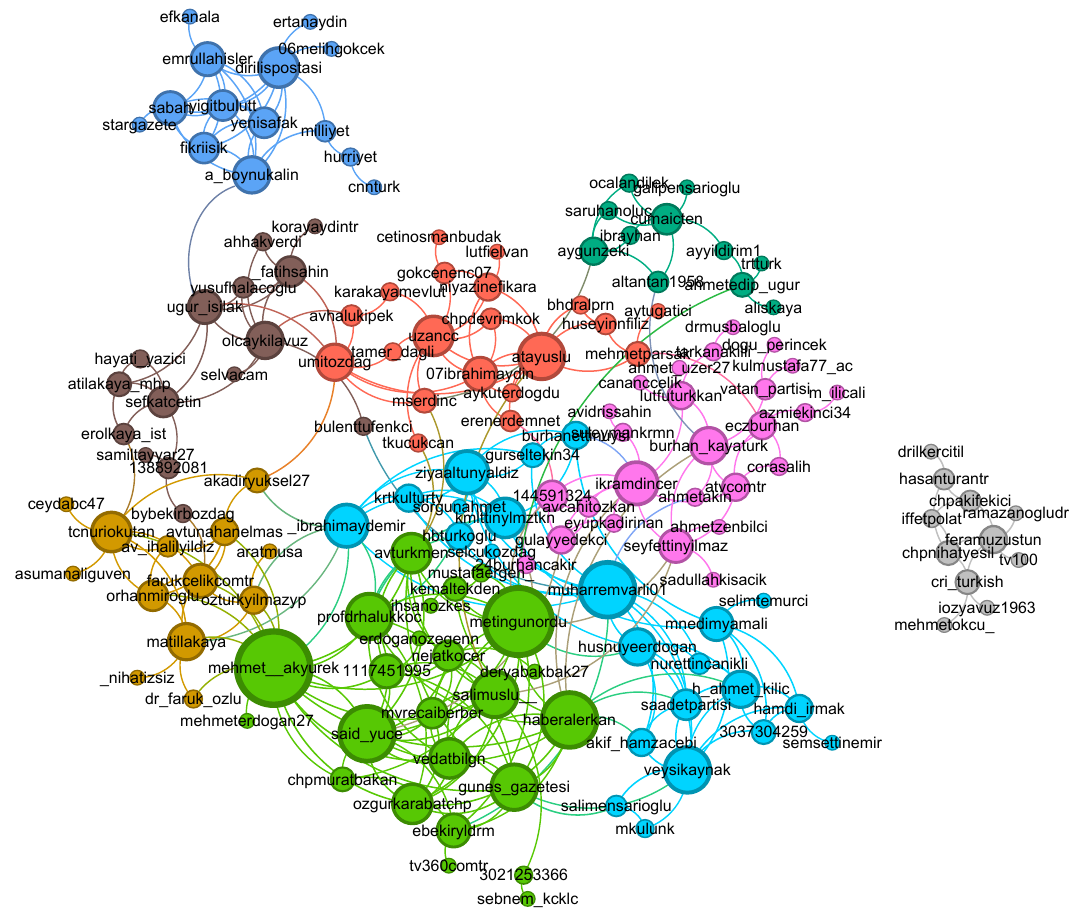}
    \caption{\textbf{Network of shared anomalous followers} Node colors represent community membership and node sizes are scaled by node degrees}
    \label{fig:full_network}
\end{figure}

\begin{figure}[H]
    \centering
    \includegraphics[width=\textwidth]{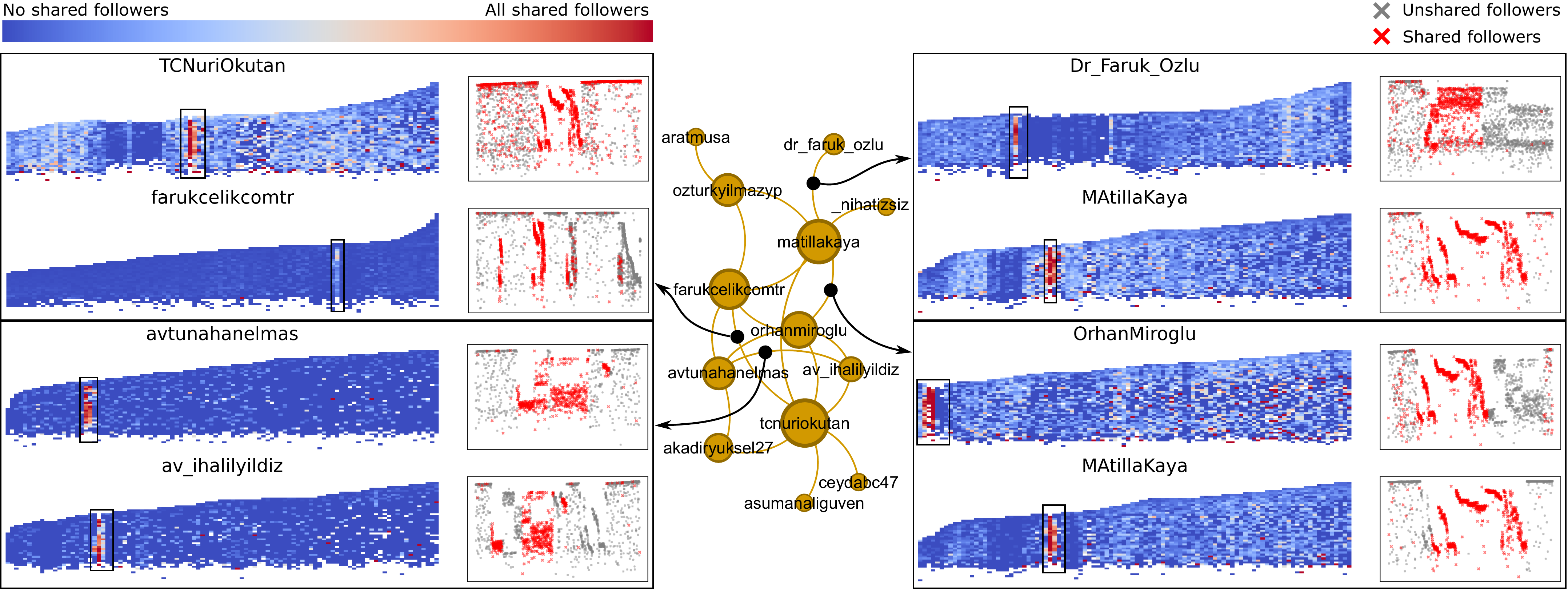}
    \caption{\textbf{Batches of anomalous followers shared across several accounts} Follower map samples of four edges in one of the communities in the user similarity network}
    \label{fig:community_sample}
\end{figure}

We create a user similarity network between the 1318 Twitter accounts in our dataset to observe the shared batches of anomalous followers between different users. The edge weight between each pair of accounts is the cosine similarity of the anomaly score vectors of the shared followers as computed from each of the follower maps of the pair of users. The six pairs of accounts corresponding to the highest six similarity scores are shown in Fig. \ref{fig:top_similarity_scores}. The follower heat map colors represent the ratio of shared followers in each bin. Each pair of followers shown in Fig. \ref{fig:top_similarity_scores} share a group of followers that are concentrated in one area of the map, i.e., accounts that followed the user consecutively. The follow patterns of these batches of shared followers are anomalous as seen in the zoomed sub-figures. Fig. \ref{fig:full_network} shows the user similarity network generated by filtering out all edge weights less than 0.75 and all edges with less than 100 shared followers. Nodes are sized by their degree and colored by their community membership. The Louvain community detection algorithm was used \cite{blondel2008fast}, which is based on modularity optimization. Fig. \ref{fig:community_sample} shows the community with the third highest pairwise average anomaly score (the first two visualized in the main text) and samples of the follower maps of connected user pairs. We observe that groups of anomalous followers that exhibit the same following pattern are observed in the followers of several accounts.

\subsection{Anomalous follower profiles} \label{A:anomalous_follower_profiles}
\begin{figure}[H]
    \centering
    \includegraphics[width=\textwidth]{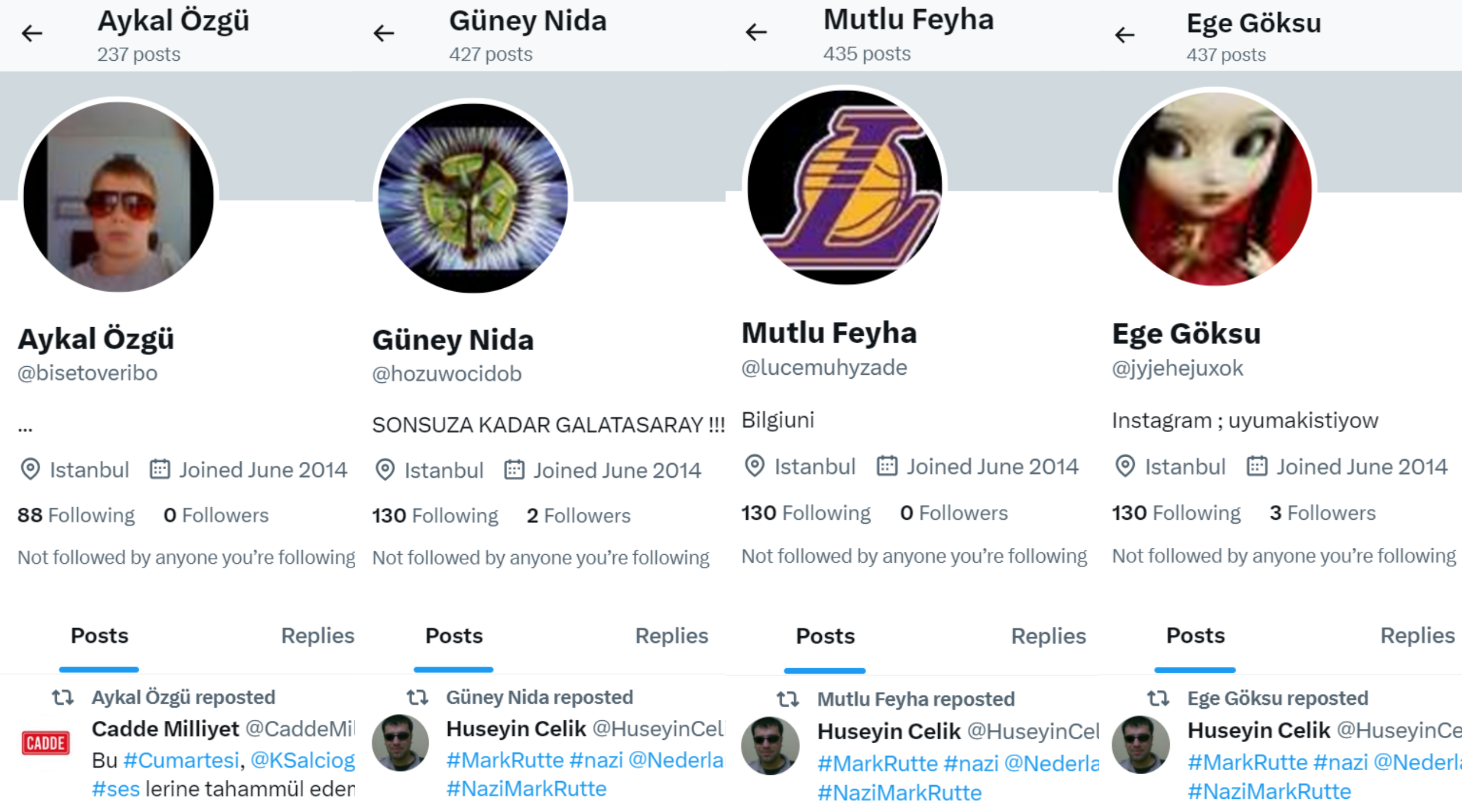}
    \caption{\textbf{Anomalous followers of @yigitbulutt} Followers have random usernames and share similar tweets. The follower map of @yigitbulut is shown in Fig. \ref{fig:anomaly_and_bot_scores}e}
    \label{fig:yigitbulutt_follower_profiles}
\end{figure}

\begin{figure}[H]
    \centering
    \includegraphics[width=\textwidth]{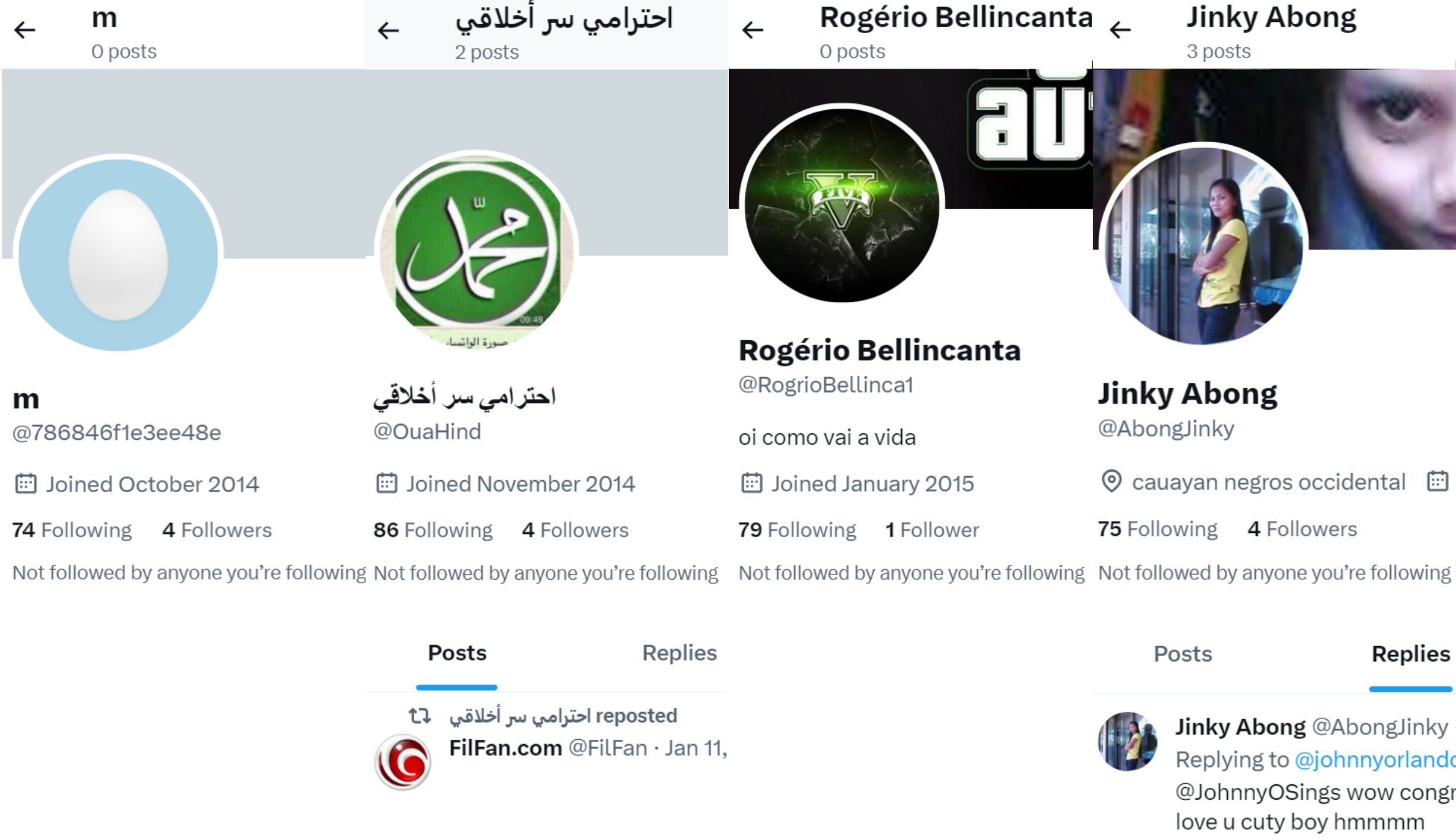}
    \caption{\textbf{Anomalous followers of @nurettincanikli} Followers impersonate individuals from different backgrounds. The follower map of @nurettincanikli is shown in Fig. \ref{fig:anomaly_and_bot_scores}a}
    \label{fig:nurettincanikli_follower_profiles}
\end{figure}

\begin{figure}[H]
    \centering
    \includegraphics[width=\textwidth]{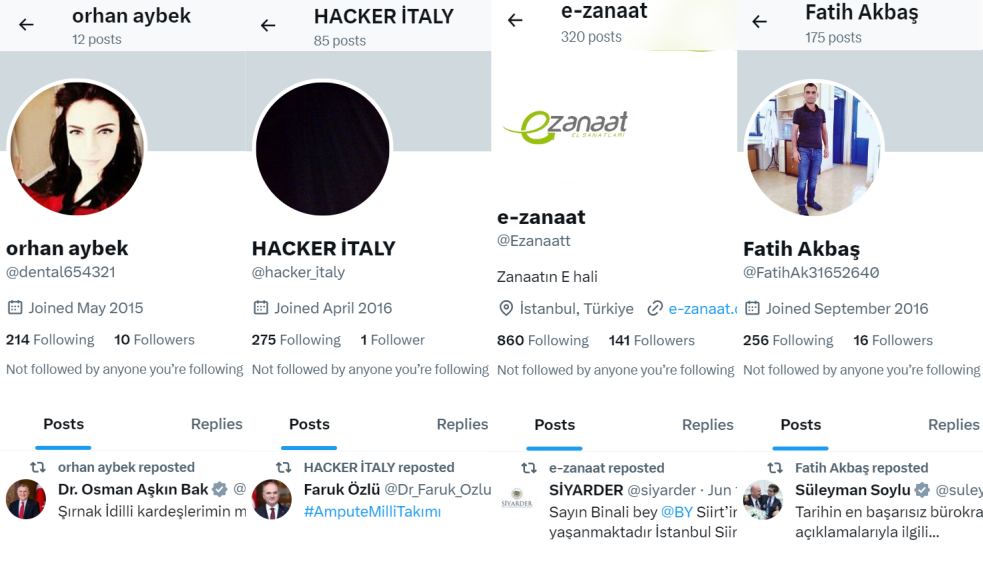}
    \caption{\textbf{Anomalous followers of @matillakaya} The follower map of @matillakaya is shown in Fig. \ref{fig:community_sample}}
    \label{fig:matillakaya_follower_profiles}
\end{figure}

Here we present samples of anomalous follower profiles. Screenshots of these profiles are shown in Fig. \ref{fig:yigitbulutt_follower_profiles}-\ref{fig:matillakaya_follower_profiles}. We also provide Internet Archive Wayback Machine snapshots documenting these profiles in Table \ref{tab:wayback_links}.

\begin{table}[H]
    \centering
    \begin{tabular}{ll}
    \hline
    Username & Anomalous follower Wayback Machine link \\
    \hline
    
    @yigitbulutt & web.archive.org/web/https://twitter.com/bisetoveribo \\
     & web.archive.org/web/https://twitter.com/hozuwocidob  \\
     & web.archive.org/web/https://twitter.com/lucemuhyzade \\
     & web.archive.org/web/https://twitter.com/jyjehejuxok \\
    \hline
    
    @nurettincanikli & web.archive.org/web/https://twitter.com/786846f1e3ee48e \\
     & web.archive.org/web/https://twitter.com/OuaHind \\
     & web.archive.org/web/https://twitter.com/RogrioBellinca1 \\
     & web.archive.org/web/https://twitter.com/AbongJinky \\
    \hline
    
    @matillakaya & web.archive.org/web/https://twitter.com/dental654321 \\
     & web.archive.org/web/https://twitter.com/hacker\_italy \\
     & web.archive.org/web/https://twitter.com/Ezanaatt \\
     & web.archive.org/web/https://twitter.com/FatihAk31652640 \\
    \hline
    \end{tabular}
    \caption{Internet Archive Wayback Machine links to the anomalous follower profiles presented in Fig. \ref{fig:yigitbulutt_follower_profiles}-\ref{fig:matillakaya_follower_profiles}.}
    \label{tab:wayback_links}
\end{table}


\end{document}